\documentclass[varenna]{cimento}

\usepackage{graphicx}  
\usepackage{amsmath}

\title{Cosmic-ray propagation in extragalactic space and secondary messengers}
\author{D.~Boncioli}
\institute{Università degli Studi dell'Aquila, Dipartimento di Scienze Fisiche e Chimiche, via Vetoio, L'Aquila, Italy\\
Istituto Nazionale di Fisica Nucleare, Laboratori Nazionali del Gran Sasso, Assergi (AQ), Italy}

\begin{document}

\maketitle

\begin{abstract}
These notes summarize the lectures about "Cosmic-ray propagation in extragalactic space and secondary messengers", focusing in particular on the interactions of cosmic-ray particles with the background photons in the Universe, including nuclear species heavier than hydrogen, and on the analytical computation of the expected cosmic-ray fluxes at Earth. The lectures were held at the Course 208 of the International School of Physics "Enrico Fermi" on "Foundations of Cosmic-Ray Astrophysics", in Varenna (Como, Italy) from June 23rd to June 29th, 2022. These notes are complementary to the content of the lectures held by Pasquale Dario Serpico at the same school.
\end{abstract}

\section{Introduction}\label{sec:intro}
\begin{figure}[t]
\centering
\includegraphics[scale=0.4]{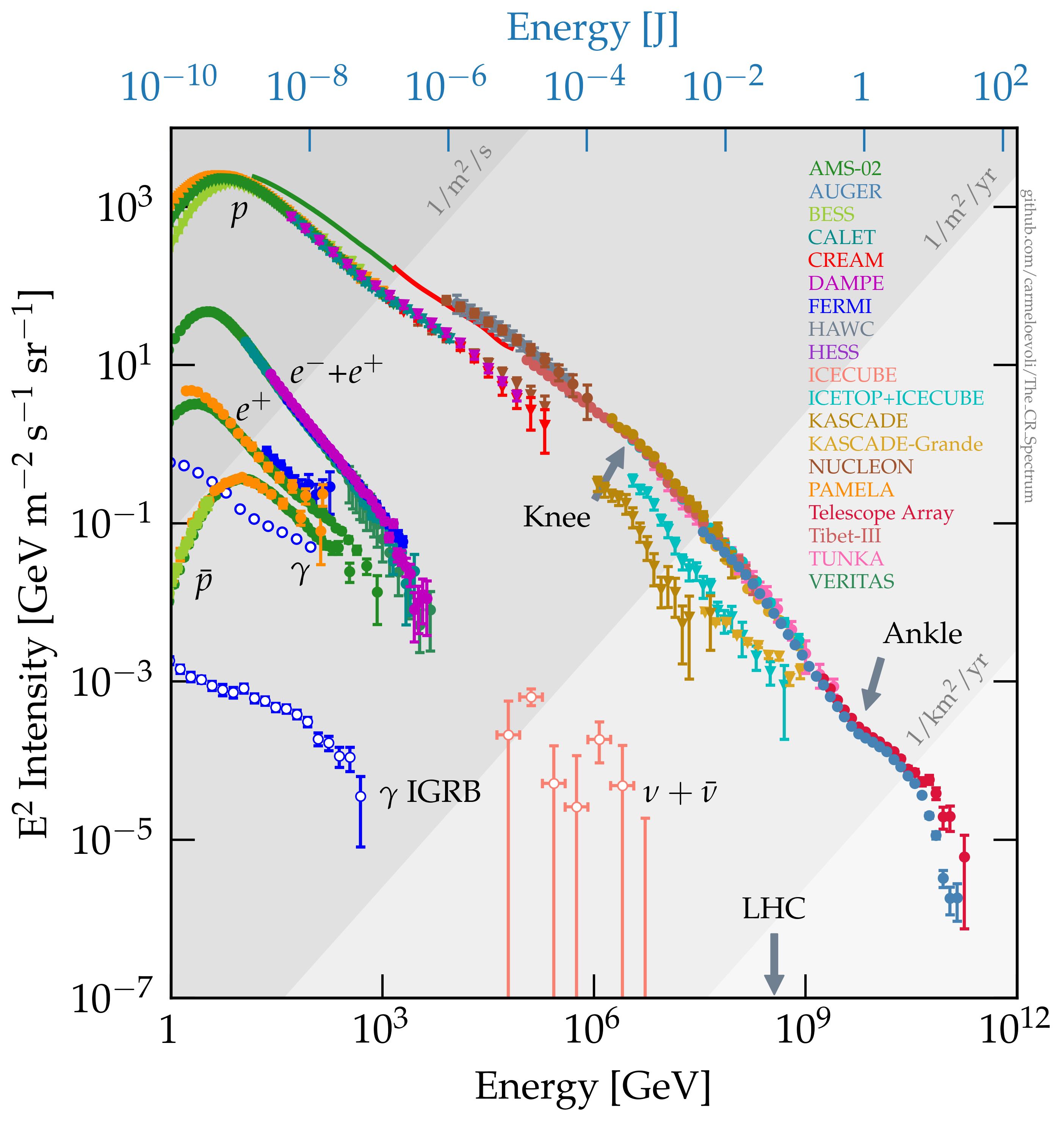}
\caption{Global view of the cosmic-ray energy spectrum, with the measurements of different experiments. From \cite{carmelo}.}
\label{fig:THEspectrum}
\end{figure}
This series of lectures focuses on the foundations of cosmic-ray astrophysics. The history of cosmic rays starts at the beginning of the past century, thanks to some experiments realized by Domenico Pacini \cite{Pacini:1912jqn} and Victor F. Hess \cite{Hess:1912srp}. They revealed that a spontaneous "radiation" was coming from the outer space, and not from the Earth crust, as it was believed to justify the spontaneous discharge of charged electroscopes. Forthcoming experiments established that this "radiation" is instead of particle nature: \textit{cosmic rays} are ionized nuclei, of which $90\%$ are protons. Thanks to cosmic rays, many fundamental discoveries in particle physics were possible until the 1950s, when high-energy particles started to be produced in human-made particle accelerators. In recent decades, cosmic rays have again been considered as the frontier to explore interactions at the highest energies, far beyond those available at terrestrial accelerators. These lectures deal with interactions of cosmic rays in astrophysical environments, whose knowledge is needed in order to unveil the astrophysical origin of cosmic rays. Beyond this aspect, cosmic rays constitute in general a unique field of research to improve our understanding of the physics governing interactions at extremely high energies. In addition, cosmic rays at the highest energies could provide a unique test of physics beyond the Standard Model, due to possible connections with dark matter as well as to probes of fundamental symmetries \cite{Coleman:2022abf}. 

Cosmic rays are measured within a vast energy range; their energy density at Earth is recognized to follow a power law spectrum as $E^{-\gamma}$ with $\gamma\sim 3$ (see Fig.~\ref{fig:THEspectrum}). The present lectures focus on the extremely high energies, namely above $10^{17}$ eV, where the cosmic rays are called Ultra-High-Energy Cosmic Rays (UHECRs). They are currently investigated thanks to giant observatories such as the Pierre Auger Observatory \cite{PierreAuger:2015eyc} and the Telescope Array \cite{TelescopeArray:2012uws}, which can measure the energy deposited in the atmosphere by the cascade of particles originated after the first interaction of the cosmic-ray particle in the atmosphere, as well as the particles reaching the surface of the Earth. Thanks to hybrid techniques, several fundamental quantities of the primary cosmic ray hitting the atmosphere can be deduced, such as its energy, arrival direction and chemical composition. UHECR particles have energy roughly more than eight orders of magnitude larger than the energy of a proton at rest, and the particles are therefore ultrarelativistic.
Although theoretical studies and experimental efforts have been developed in the last decades, several issues concerning UHECRs are still unsolved. The astrophysical origin of cosmic rays, as well as their chemical composition and the mechanisms that bring them to such extreme energies in their sites of production, are uncertain, and constitute a major and exciting field of study in astroparticle physics. 

\begin{figure}[t]
\centering
\hspace{-0.2cm}
\begin{tabular}{cc}
\includegraphics[scale=0.32]{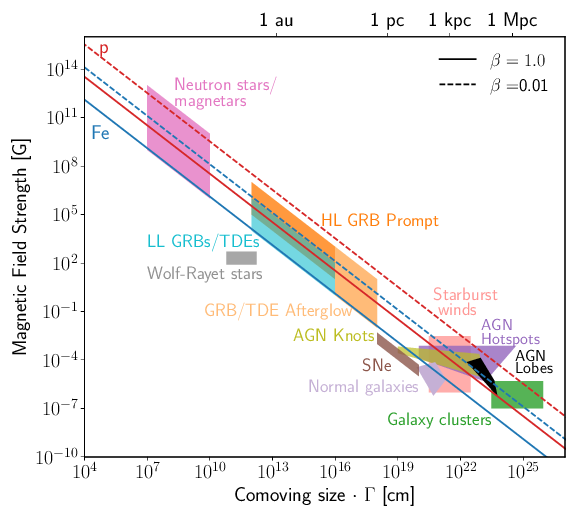}&
\includegraphics[scale=0.31]{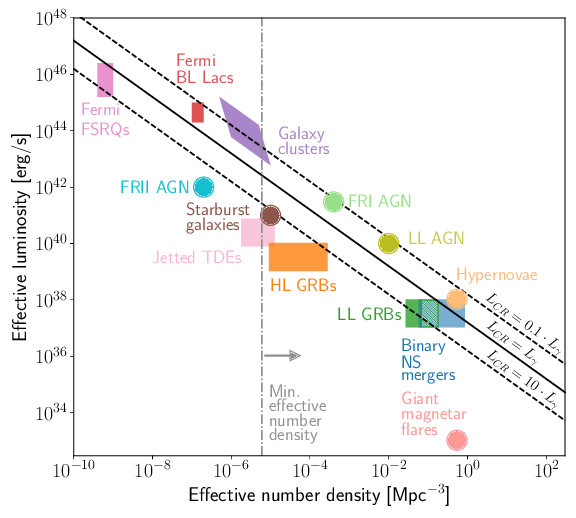}
\end{tabular}
\caption{Left: Source classes as a function of their comoving size and magnetic field, and corresponding maximum energy reachable for hydrogen and iron nuclei for different values of the shock velocity. Right: Characteristic source luminosity versus source number density for steady sources, and effective luminosity versus effective number density for transient sources. Both figures are reproduced with permission from \cite{AlvesBatista:2019tlv}.}
\label{fig:hillas}
\end{figure}

Due to the extremely high energy of UHECRs, and taking into account the confinement power of the magnetic fields in the Galaxy, they are not expected to be produced in the Galaxy, as also significantly confirmed by large-scale anisotropy studies \cite{PierreAuger:2017pzq}. In order to estimate what type of extragalactic sources can host mechanisms to accelerate particles to such high energies, the acceleration mechanisms of particles in Supernova Remnants (SNRs), as predicted in the standard paradigm of the Galactic cosmic rays \cite{Gabici:2019jvz}, can be exploited. In these environments, the maximum energy reachable from particles due to the presence of shocks is connected to the age, and therefore to the dimension, of the SNR, as well as to the intensity of the magnetic fields \cite{Lagage:1983zz,Drury:1983zz}. This has been generalized in \cite{Hillas:1984ijl}, and the maximum energy can be thus defined in terms of the confinement power of the astrophysical source, meaning that the particles can reside in the acceleration region as soon as the gyroradius is smaller than the region itself. This argument, known as "Hillas condition", permits to classify the candidate sources in terms of their comoving size and magnetic field in the acceleration region, as shown in Fig.~\ref{fig:hillas} (left), where the maximum energy is expressed as $E_{\max}=Ze \beta_{\mathrm{sh}} c BR$, being $\beta_{\mathrm{sh}}$ the velocity of the shock in units of speed of light, $c$, $B$ the intensity of the magnetic field in the source, $R$ the size of the accelerating region and $Ze$ the charge of the particle. Several source types reported in Fig.~\ref{fig:hillas} (left) are capable of accelerating cosmic rays to ultra-high energies. A complementary condition to be taken into account is that the considered source class must produce the energy budget in cosmic rays to account for the observed energy flux at Earth. From the fit of the UHECR spectrum and composition measured at Earth, an estimate of the UHECR energy production rate per unit volume (also called luminosity density or emissivity, being the product of the luminosity and the number density) can be given, as done in \cite{PierreAuger:2016use}, where $5\times 10^{44}$ $\mathrm{erg \, Mpc^3 \, yr^{-1}}$ is found. Fig.~\ref{fig:hillas} (right) shows what source classes satisfy this requirement, taking into account the combination of luminosity and number density. The vertical line in the plot highlights the minimum effective number density that can be estimated from studies involving arrival directions. The full diagonal line refers to the case in which the luminosity in cosmic rays is supposed to coincide with the gamma-ray luminosity. The plots in Fig.~\ref{fig:hillas} summarize generic indications about source classes that can be considered as responsible for the UHECR flux at Earth. More detailed indications come, for instance, from studies that compare UHECR arrival directions with the position of sources from catalogs, as in \cite{PierreAuger:2018qvk} where the starburst galaxies are found to be correlated with the highest energy CRs at more than 4.0 $\sigma$. Studies that interpret the UHECR energy spectrum and composition in terms of astrophysical scenarios can in turn constrain the spectral characteristics of UHECRs at the escape from their sources \cite{PierreAuger:2016use}.

Some slight departures from a single power law in the CR spectrum can be recognized; for instance, a change of slope is visible at $\sim 3\times 10^{15}$ eV, called \textit{knee}. This could be interpreted as a signature of the end of the acceleration power of CR sources at work in the Galaxy, as due to processes in SNRs. Another change of slope, measured at $\sim 5\times 10^{18}$ eV, called \textit{ankle}, could be connected to the intersection of the spectra of Galactic and extragalactic CRs, as well as to different contributions from populations of extragalactic sources, or to effects of the energy losses of CRs during their travel through the extragalactic space (see for instance Ref.~\cite{PierreAuger:2022atd} and references therein). The origin of the suppression of the CR spectrum at the highest energies is also undetermined, as it will be also mentioned in the following. For a comprehensive and recent review of the open questions about cosmic rays, see Ref.~\cite{AlvesBatista:2019tlv}.\\

The lecture notes are organized as follows. In Sec.~\ref{sec:rate} I will introduce the physics of interactions that take place in the travel of the UHECR particles in the space they traverse before being detected. Sec.~\ref{sec:UHECRfluxes} is dedicated to the analytical computation of the cosmic-ray flux at Earth, for the case of protons and heavier nuclei, respectively. In this section, several references to the \textit{SimProp} Monte Carlo code for the UHECR extragalactic propagation \cite{Aloisio:2017iyh}, of which I am one of the authors, can be found, as well as to some applications of the code, that is currently under revision for improving its performances and for refining some physics inputs \cite{Sirente}.  In Sec.~\ref{sec:neutrinos} the production of secondary particles that are expected to be generated by cosmic rays interacting in the extragalactic background photon field is described. Several appendices complement the material reported in the main text.
\section{Interactions of UHECRs}\label{sec:rate}
 In this section the calculation of the energy losses of cosmic-ray protons as due to interactions with the cosmic microwave background (CMB) is worked out analytically; the same calculation can be performed for the case of a generic photon field, taking into account its dependence on the redshift and energy. The case of nuclei heavier than hydrogen will be also shown.

As a first step, the available energy for the interactions is discussed; due to the relativistic energies of the involved particles, in order to describe a process such as $a+b \rightarrow c+d$, where $a,\,b,\,c,\,d$ are the particles in the initial and final states,  their energy-momentum four-vectors can be defined as 
\begin{equation}
 p_{\mu} = (E_i,\,c\vec{p}_i);
    \label{eq:fourvec}
\end{equation}
another quantity to be considered is the $s$ of the process, namely the scalar product of the cumulative four-vectors of the initial and final states, that is a Lorentz-invariant quantity. As a general approach applied to any of the processes described in the following, the value of $s_{\mathrm{th}}$, namely the $s$ at the threshold for the reaction, will be computed in a convenient reference frame, such as
\begin{equation}
s_{\mathrm{th}}=(E_a+E_b)^{2}-c^{2}(\vec{p}_a+\vec{p}_b)^{2}=(m^{2}_c+m^{2}_d)c^{4},
    \label{eq:sth_def}
\end{equation}
corresponding to the laboratory frame. This quantity will be calculated corresponding to the interactions relevant in the propagation of the particles through the extragalactic space. 

The necessary ingredients for computing the rates of the processes are the spectra of the photons encountered during the extragalactic propagation; the ones relevant for UHECRs are mainly the CMB and the ultraviolet-optical-infrared background light (also called Extragalactic Background light, EBL). The cross sections of the photo-nuclear processes, that will be discussed in the following, have to be included in the computation as well.\\

\begin{figure}[t]
\centering
\includegraphics[scale=0.35]{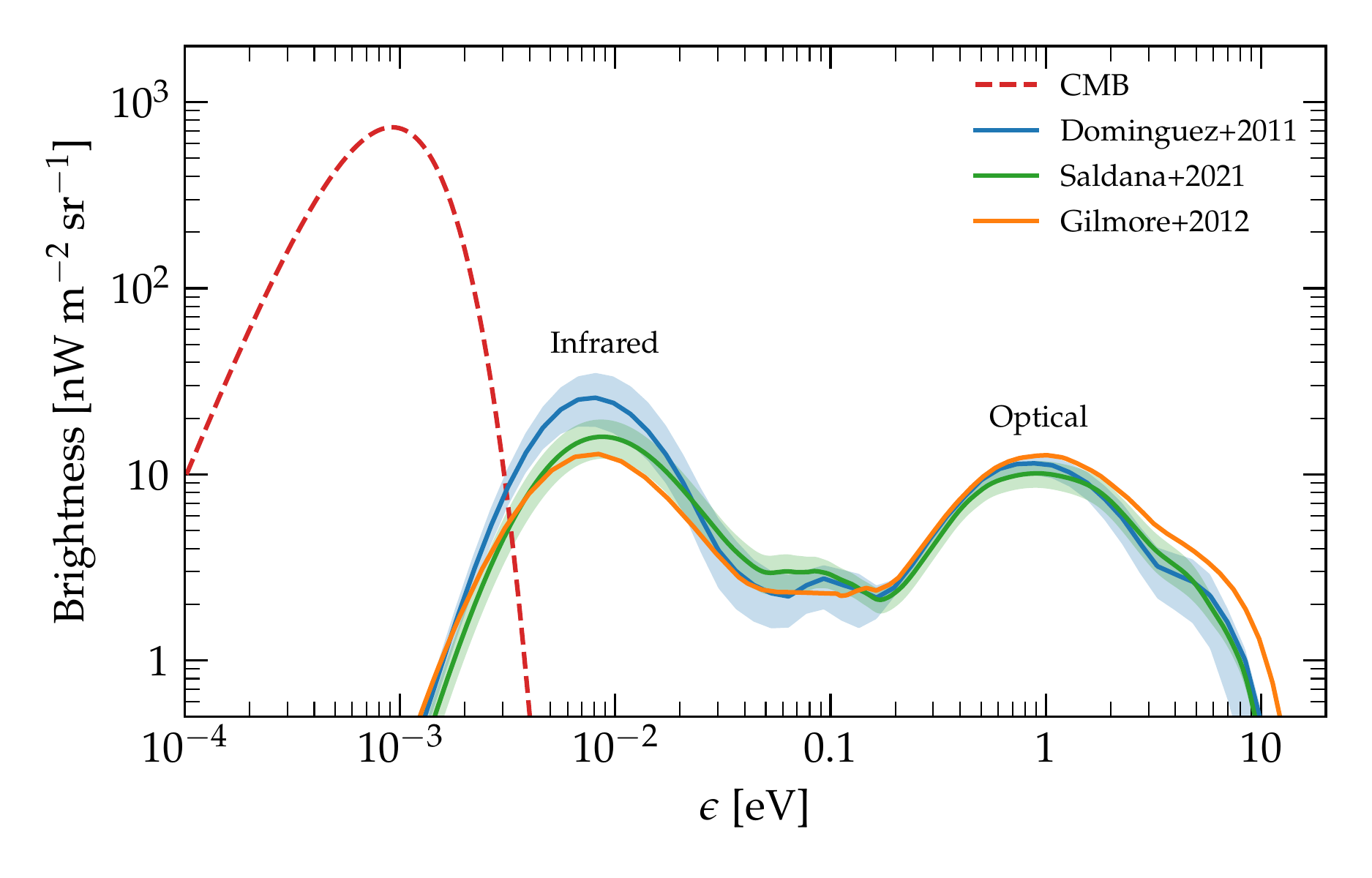}
\caption{Spectral energy density of the CMB and EBL calculated in the local Universe, according to the models reported in \cite{Dominguez:2010bv,Gilmore:2011ks,Saldana-Lopez:2020qzx}, and used in \cite{Sirente}.}
\label{fig:CMB_EBL}
\end{figure}

\begin{figure}[h]
\centering
\includegraphics[scale=0.35]{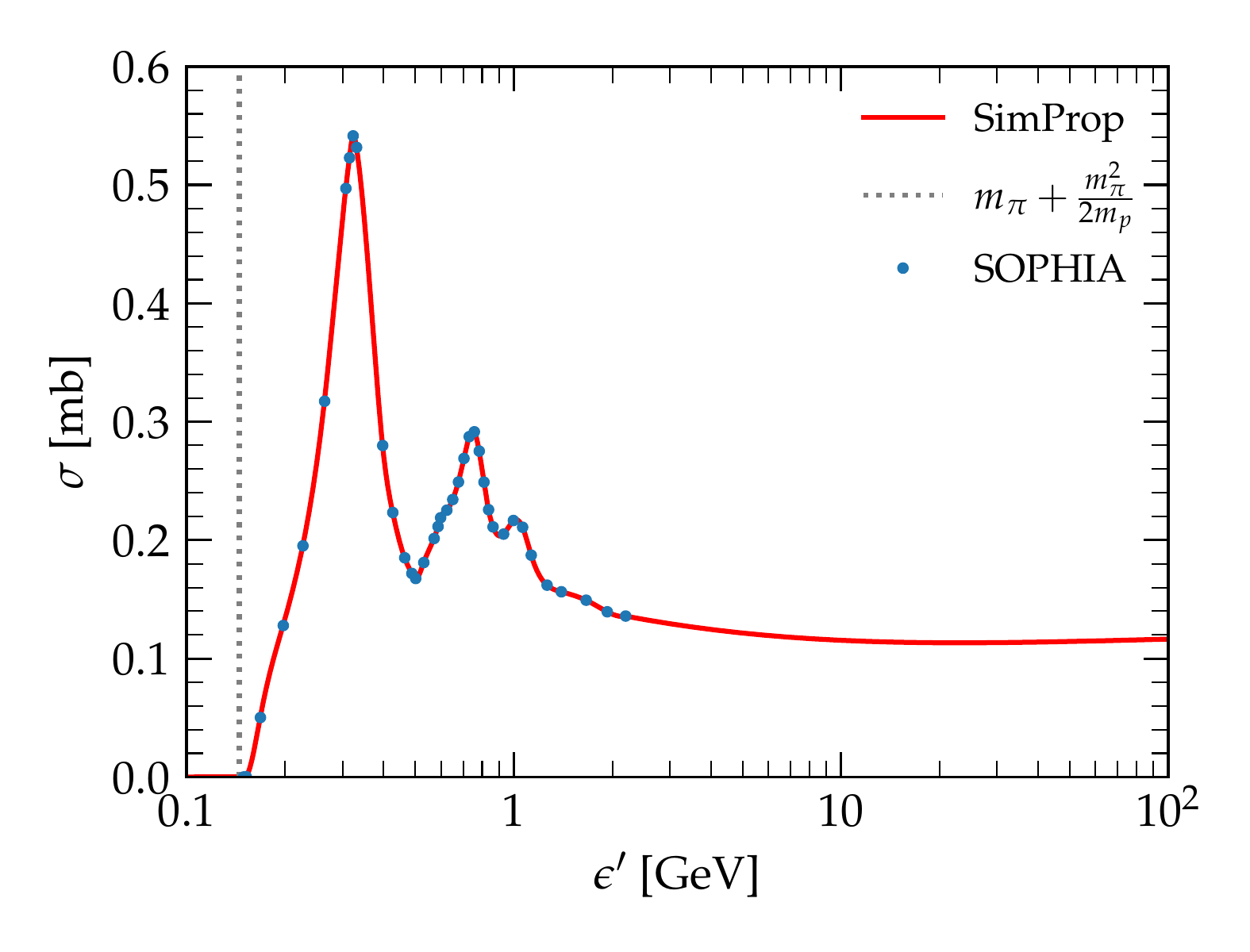}
\caption{Total cross section for the photo-meson production process as a function of the energy of the photon in the proton rest frame; the cross section is taken from SOPHIA \cite{Mucke:1999yb} and used in \cite{Sirente}.}
\label{fig:xsection_p}
\end{figure}

The typical time of an interaction process is approximately proportional to $1/(c \sigma n_{\mathrm{ph}})$, where $\sigma$ represents the cross section of the process, while $n_{\mathrm{ph}}$ is the density of the photons encountered by the cosmic ray in the extragalactic space. If the distribution of these quantities in terms of energies is considered, the interaction rate can be computed as:
\begin{equation}
\frac{dN_{\mathrm{int}}}{dt} = c\int(1-\beta \cos\theta) n_{\mathrm{ph}}(\epsilon,\cos\theta) \sigma(\epsilon') d\cos\theta d\epsilon \, ,
\label{eq:intl} 
\end{equation}
where $n_{\mathrm{ph}}$ is the energy spectrum of the photon field (as a function of the energy of the photon in the laboratory frame and of the angle between the momenta of the particle and the photon) and $\sigma$ is the cross section of the considered process, as a function of the energy of the photon in the particle frame (here and in the following the primed terms refer to the quantities computed in the proton/nucleus rest frame). These quantities are reported respectively in Fig.~\ref{fig:CMB_EBL} (calculated in the local Universe, for the CMB and some models of the EBL) and Fig.~\ref{fig:xsection_p}.\\

From the energy of the photon in the proton rest frame, the transformation $d\epsilon'=-\Gamma \epsilon d \cos\theta$ is derived and the integral reads then as two integrals over the energy of the photons (whose distribution is assumed isotropic in the laboratory frame), in the laboratory and the proton rest frame: 
\begin{equation}
\frac{dN_{\mathrm{int}}}{dt} = \frac{c}{2\Gamma^{2}}\int^{\infty}_{\epsilon'_{\mathrm{th}}} \sigma(\epsilon') \epsilon' \int^{\infty}_{\epsilon'/2\Gamma} \frac{n_{\mathrm{ph}}(\epsilon)}{\epsilon^{2}}   d\epsilon d\epsilon' \, .
\label{eq:intl_general} 
\end{equation}
A complete derivation of the interaction rate is reported in App.~\ref{sec:interactionrate}.\\

In order to compute the energy loss length, the inelasticity of the processes taken into account has to be evaluated, meaning the mean fraction of energy of a nucleus lost in a single interaction, $f(\epsilon')=\langle \frac{E_{\mathrm{in}}-E_{\mathrm{out}}}{E_{\mathrm{in}}}\rangle$. The interaction length in Eq.~\ref{eq:intl_general} can be used to compute the energy loss rate, by introducing the inelasticity, as:
\begin{equation}
\frac{1}{E}\frac{dE}{dt} = - \frac{c}{2\Gamma^{2}}\int^{\infty}_{\epsilon'_{\mathrm{th}}} f(\epsilon') \sigma(\epsilon') \epsilon' \int^{\infty}_{\epsilon'/2\Gamma} \frac{n_{\mathrm{ph}}(\epsilon)}{\epsilon^{2}}   d\epsilon d\epsilon' \, ,
\label{eq:elr_general} 
\end{equation}
for a generic process. This rate can be thus converted into a length as:
\begin{equation}
l_{\mathrm{loss}} = -c \left(\frac{1}{E}\frac{dE}{dt} \right)^{-1} = -E \frac{dx}{dE}
\label{eq:ell_general} 
\end{equation}
and used to follow the trajectory of the particle as
\begin{equation}
\frac{dE}{dx} = -\frac{E}{l_{\mathrm{loss}}}
\label{eq:trajectory} 
\end{equation}
being $x$ the distance covered by the particle.\\

Let us calculate the interaction rate in the case of UHECR protons interacting with the CMB. Already at the time of the discovery of the CMB, it was supposed that the photo-pion production $p+\gamma_{\mathrm{bkg}} \rightarrow p(n) + \pi^{0}(\pi^{+})$ of protons off CMB photons could cause energy losses inducing a suppression of the UHECR flux \cite{Greisen:1966jv,Zatsepin:1966jv}. The threshold for this process can be calculated as:
\begin{equation}
   s_{\mathrm{th}}=m^{2}_{\mathrm{p}}c^{4}+2E_{\mathrm{th}}\epsilon(1-\beta \cos \theta)=(m_{\mathrm{p}}+m_{\mathrm{\pi}})^{2}c^{4}\,
\label{eq:sth_pion}    
\end{equation}
where $m_{\mathrm{p}},m_{\mathrm{\pi}}$ are the proton and the pion masses, respectively, $\beta$ is the speed of the proton in the laboratory frame (being the particles ultrarelativistic, this can be taken as $\beta\sim 1$ and will be omitted in the following), $\theta$ is the angle between the photon and the proton momenta, $\epsilon$ is the energy of the photon in the laboratory frame and $E_{\mathrm{th}}$ is the minimum energy required for the proton in order to induce a pion production, which reads:
\begin{equation}
   E^{\pi}_{\mathrm{th,p}}=\frac{m^{2}_{\mathrm{\pi}}c^{4}+2m_{\mathrm{\pi}}m_{\mathrm{p}}c^{4}}{2\epsilon(1-\cos\theta)}\approx 7\times10^{19}\, \mathrm{eV}\,
\label{eq:sth_pion}    
\end{equation}
if head-on collisions are taken into account with a photon of $\epsilon \approx 7 \times 10^{-4}\,\mathrm{eV}$ (average CMB photon energy). 

Another process that can cause energy losses of CR protons is the electron-positron pair production \cite{Blumenthal:1970nn}, $p+\gamma_{\mathrm{bkg}} \rightarrow p + e^{+}+e^{-}$, for which the energy threshold can be calculated similarly to the case of the pion production:
\begin{equation}
   E^{e^{+}e^{-}}_{\mathrm{th,p}}=\frac{4m^{2}_{\mathrm{e}}c^{4}+8m_{\mathrm{e}}m_{\mathrm{p}}c^{4}}{2\epsilon(1-\cos\theta)}\approx 6\times10^{17}\, \mathrm{eV}\,.
\label{eq:sth_pair}    
\end{equation}
In order to evaluate what photon fields can play a role in these processes, one can compute the energy of the photon in the proton rest frame: $\epsilon'=\epsilon\Gamma(1-\cos\theta)\approx   \epsilon\Gamma$. Therefore the needed threshold Lorentz factor to trigger a photo-pion production in the EBL (mean infrared energy $10^{-1}\div 10^{-2}$ eV) will be lower than what is found for the CMB photons, therefore permitting lower energy protons to induce the production of pions. Although the pion production by protons off EBL is less efficient in terms of energy losses if compared to the pair production in the same energy range, this is relevant for the production of neutrinos (as discussed in Sec.~\ref{sec:neutrinos}).

If the energy spectrum of CMB photons (black body)
\begin{equation}
    n_{\mathrm{ph}} = \frac{dN_{\mathrm{ph}}}{dV d\epsilon} = \frac{1}{\pi^{2} (\hbar c)^{3}} \frac{\epsilon^{2}}{\exp(\epsilon/k_{\mathrm{B}}T)-1}
    \label{eq:n_CMB} 
\end{equation}
is considered (where isotropy is also assumed, so that $n_{\mathrm{ph}}(\epsilon,\cos\theta)\approx n_{\mathrm{ph}}(\epsilon)$), the calculation of the integral over the photon density in Eq.~\ref{eq:intl_general} can be worked out analytically \cite{Berezinsky:2002nc}, with the transformation $y=\exp(\epsilon/k_{\mathrm{B}}T)-1$. The interaction rate becomes then
\begin{equation}
\frac{dN_{\mathrm{int}}}{dt} = \frac{c k_{\mathrm{B}}T}{2\pi^{2}(\hbar c)^{3}\Gamma^{2}}\int^{\infty}_{\epsilon'_{\mathrm{th}}} \sigma(\epsilon') \epsilon' \left\{ -\ln \left[ 1-\exp \left( -\frac{\epsilon'}{2\Gamma k_{\mathrm{B}}T}\right) \right] \right\} d\epsilon' \, .
\label{eq:intl_CMB} 
\end{equation}

The inelasticity at the threshold for the production can be computed taking into account the masses of the particles to be generated, so that  $f_{\pi}(\epsilon'\approx 145\, \mathrm{MeV})=\frac{m_{\mathrm{\pi^{0}}}}{m_{\mathrm{p}}+m_{\mathrm{\pi^{0}}}}\approx 0.125$ and $f_{e^{+}e^{-}}(\epsilon'\approx 1\, \mathrm{MeV})=\frac{2m_{\mathrm{e}}}{m_{\mathrm{p}}+2m_{\mathrm{e}}}\approx 10^{-3}$ corresponding to photo-pion and pair production respectively. In order to compute the same quantities at energies higher than the threshold, the particle distributions in the final states are needed. In the energy range of interest for CR interactions, the inelasticity is usually approximated as 20\% for the photo-pion production and $10^{-5}$ for the pair production \cite{Gaisser2}.

\begin{figure}[t]
\centering
\includegraphics[scale=0.35]{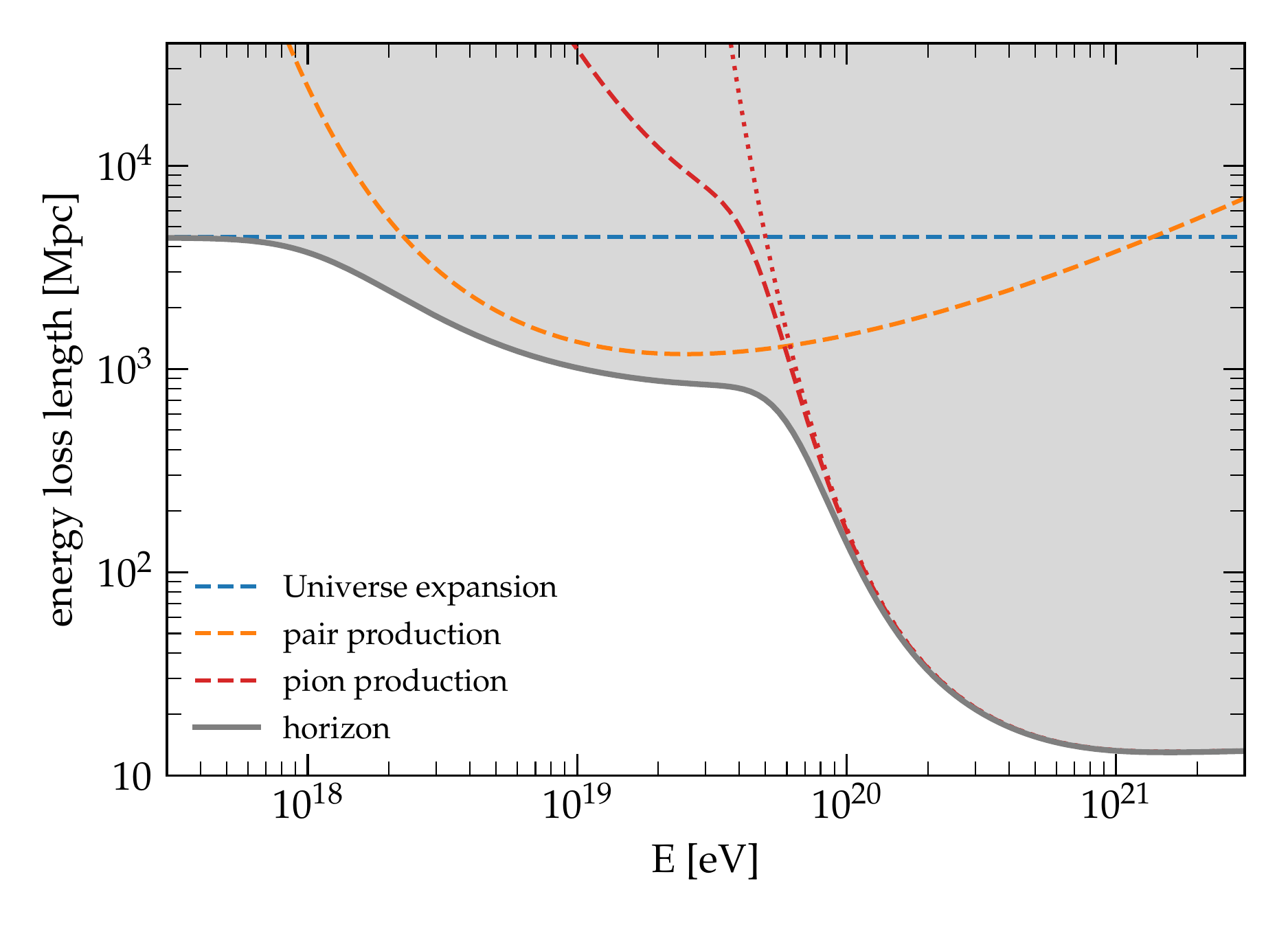}
\caption{Total energy loss length for protons (solid line) as a function of the Lorentz factor, calculated at redshift $z=0$, with the contributions from photo-pion production (red-dashed line refers to the total photo-pion energy loss length while the red-dotted line refers only to the photo-pion energy loss length off CMB) and from electron-positron pair production (orange dashed line) off CMB photons \cite{Sirente}.}
\label{fig:lambda_p}
\end{figure}

 Fig.~\ref{fig:lambda_p} shows the energy loss lengths corresponding to different processes for protons in the CMB, as a function of the energy, and computed at the present time (or redshift $z=0$), with the inelasticity taken into account for the case of isotropic production of the pions in the center of mass frame (other options are described in \cite{Hummer:2010vx}). The horizontal line shows the adiabatic energy losses due to the expansion of the Universe, corresponding to $z=0$, that are in general given by: 
\begin{equation}
\frac{1}{E}\frac{dE}{dt} = -H(z(t))\, ,
\label{eq:ell_ad} 
\end{equation}
where $H(z)=H_0 \sqrt{(1+z)^{3}\Omega_m+\Omega_{\Lambda}}$ (see also App. \ref{sec:cosmology} for more details). At intermediate energies the effect of the energy losses due to the pair production is dominant; the typical length traversed by UHECRs undergoing these processes is of the order of Gpc. At the highest energies photo-pion processes can take place, and the typical length is of the order of 10 Mpc. Taking into account the fraction of proton energy lost in each interaction, protons with initial energy of $10^{21}$ eV will be under threshold for photo-pion production after traveling $50\div 100$ Mpc. This means that if UHECRs with EeV energies are detected at Earth, these should be produced within a sphere of the order of $\sim$ 100 Mpc, as predicted by \cite{Greisen:1966jv,Zatsepin:1966jv}.

The energy loss lengths in Fig.~\ref{fig:lambda_p} are computed corresponding to the present time. Due to the dependence of the density of the CMB photons on the redshift and the dependence of the temperature of the CMB photons, the energy loss length varies as:
\begin{equation}
    l_{\mathrm{loss}}(E,z) = \frac{l_{\mathrm{loss}}((1+z)E,z=0)}{(1+z)^{3}}\, ,
\label{eq:ell_z}     
\end{equation}
while if the EBL is used instead of the CMB, the expression above has a more complicate dependence on $z$. The complete computation of Eq.~\ref{eq:ell_z} is reported in App.~\ref{sec:redshiftdependence}.\\

Current measurements of cosmic rays at the highest energies are found to be inconsistent with a pure-proton composition, using current hadronic interaction models for taking into account the development of the cascade of particles generated in the atmosphere after the first interaction of the primary cosmic ray \cite{PierreAuger:2014sui,PierreAuger:2014gko}. 

\begin{figure}[t]
\centering
\includegraphics[scale=0.3]{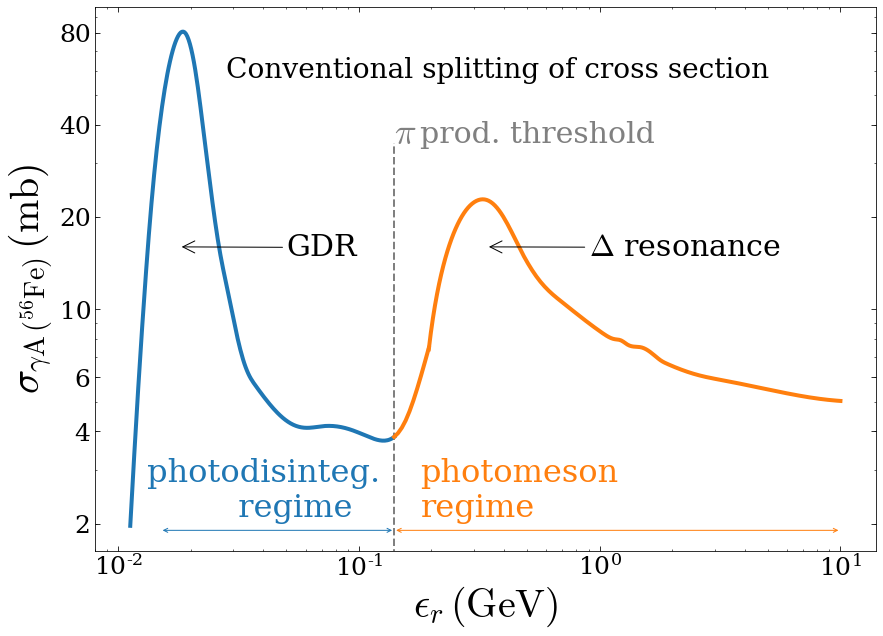}
\caption{Total inelastic photo-nuclear cross section for iron-56
 as a function of photon energy in the rest frame of the nucleus. Reproduced with permission from \cite{Morejon:2019pfu}.}
\label{fig:crosssection56}
\end{figure}

If cosmic rays reaching the top of the atmosphere are heavier than protons, their possible interactions must be taken into account for the propagation through extragalactic photons in order to possibly infer the UHECR mass composition and spectral parameters at their sources. In addition to the electron-positron pair production and the photo-pion production, the photo-disintegration process plays an important role in the modification of the nuclear species of the cosmic rays as escaped from their sources, on their way to the Earth. Unlike the pion production, the disintegration of nuclei can be triggered correspondingly to energies of the photon in the nucleus rest frame of tens of MeV, as shown for instance in Fig.~\ref{fig:crosssection56} for the case of iron-56. At these energies it is possible to neglect the binding energy of the nucleons in the nucleus, therefore the energy loss lengths can be computed as separated contributions from the modification of the Lorentz factor (due to energy losses from adiabatic expansion, pair production and pion production) and the change in the atomic mass number, due to the photo-disintegration, where the Lorentz factor is conserved \cite{Aloisio:2008pp}:
\begin{equation}
    \frac{1}{E}\frac{dE}{dt} = \frac{1}{\Gamma}\frac{d\Gamma}{dt} + \frac{1}{A}\frac{dA}{dt}\, .
\end{equation}
The photo-disintegration process comprises two main regimes \cite{Puget:1976nz,Boncioli_PhDThesis2011}, as shown in Fig.~\ref{fig:crosssection56}:
\begin{itemize}
    \item a resonance at about 10 MeV (energy of the photon in the nucleus rest frame, slightly dependent on the nucleus), called Giant Dipole Resonance (GDR), corresponding to the behavior of protons and neutrons in the nucleus as penetrating fluids; the de-excitation of this resonance produces the ejection of one or two nucleons;
    \item a flat region in the range 20 - 150 MeV, called Quasi-Deuteron regime, where the photon wavelength in the nucleus rest frame is comparable to the nuclear dimensions and the photon is likely to interact with a nucleon pair, with the final ejection of that pair and possibly other nucleons.
\end{itemize}

\begin{figure}[h]
\centering
\includegraphics[scale=0.2]{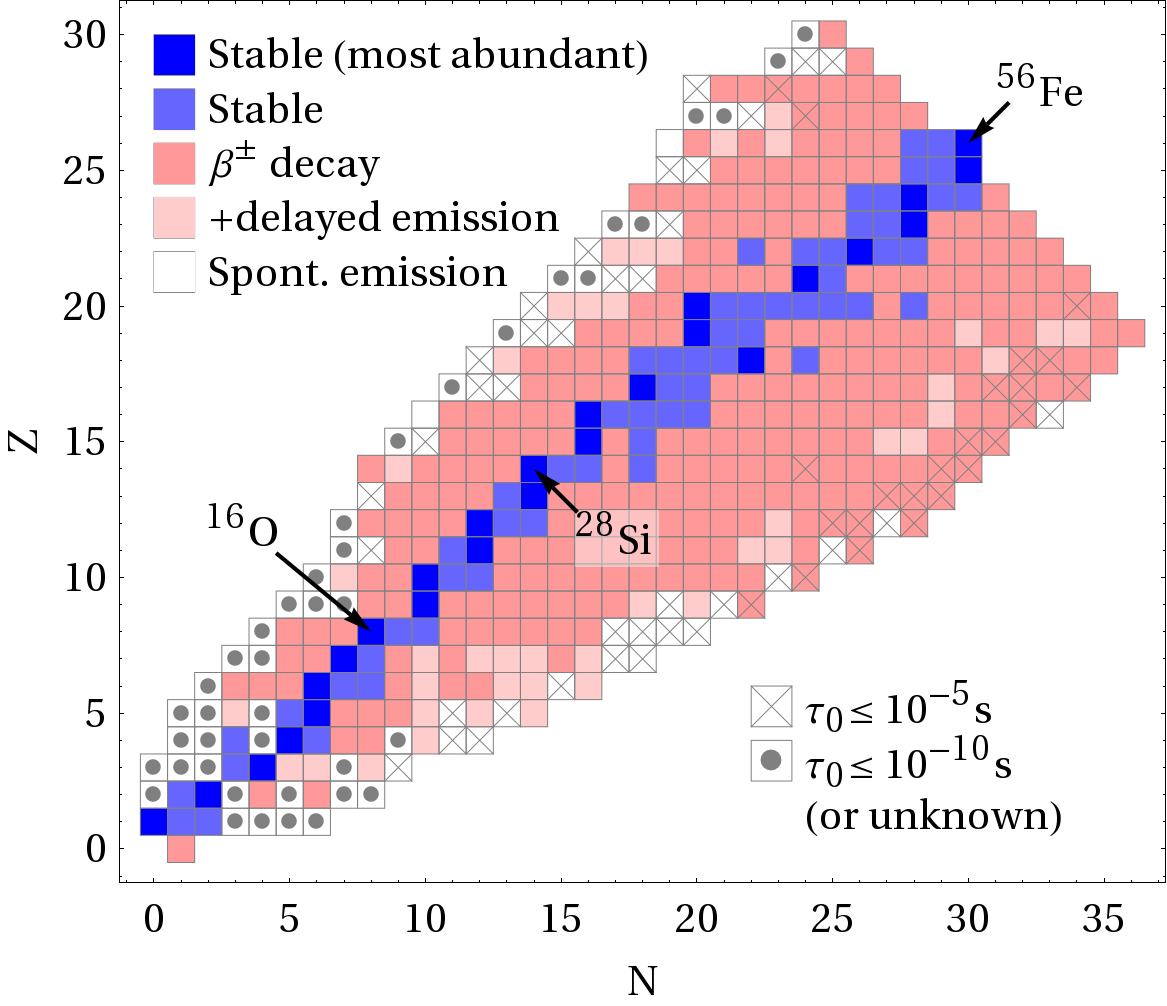}
\caption{Nuclear chart as a function of the number of protons and neutrons, showing the isotopes that can be created in the cascade after the first interaction of an isotope of iron-56 with a background photon field. Reproduced with permission from \cite{Biehl:2017zlw}.}
\label{fig:crosssection_chart}
\end{figure}

\begin{figure}[t]
\centering
\includegraphics[scale=0.45]{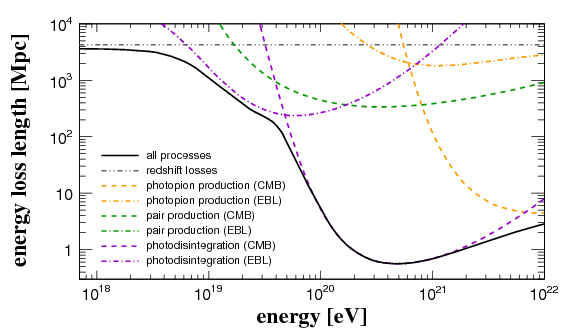}\\
\includegraphics[scale=0.45]{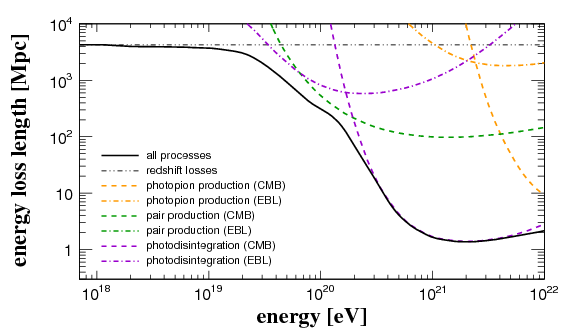}
\caption{Total energy loss lengths for nitrogen-14 (top) and iron-56 (bottom), calculated at $z=0$, and the contributions from the different processes. Reproduced with permission from \cite{AlvesBatista:2015jem}.}
\label{fig:ell_nuclei}
\end{figure}

A cascade of isotopes lighter than the injected primary can be therefore generated due to photo-disintegration in astrophysical environments, as reported in Fig.~\ref{fig:crosssection_chart} for the iron-56 as primary nucleus. 

Examples of energy loss lengths are shown in Fig.~\ref{fig:ell_nuclei}, corresponding to nitrogen-14 (top) and iron-56 (bottom) nuclei. Similarly to the case of protons, the adiabatic energy losses are dominant at low energies, and the corresponding energy loss length can be written as in Eq.~\ref{eq:ell_ad}. At intermediate energies, the pair production\footnote{The energy loss length for nuclei for pair production is corrected by a factor $Z^{2}/A$ with respect to the case of protons; here the factor $Z^{2}$ takes into account the correction due to the cross section, while the factor $1/A$ takes into account the change in the inelasticity \cite{Aloisio:2008pp}.} is overcome by the photo-disintegration on the EBL, while at the highest energies the dominant process is the photo-disintegration on the CMB. The pion production is shifted towards higher energies with respect to the case of protons. This is due to the fact that in this process the particle involved is the nucleon in the nucleus, therefore the threshold is $E_{\mathrm{th}}\approx A\Gamma^{p}_{\mathrm{th}} m_p c^{2}$, where $\Gamma^{p}_{\mathrm{th}}$ can be derived from Eq.~\ref{eq:sth_pion}. Therefore, the pion production will be more efficient in the case of protons with respect to heavier nuclear species, and this will have consequences for the production of secondary messengers (see Sec.~\ref{sec:neutrinos}).\\

\section{Computation of UHECR fluxes at Earth}\label{sec:UHECRfluxes}
The propagation of UHECRs in the intergalactic or Galactic medium can be followed with a system of differential equations describing the evolution of the particles with respect to the time, taking into account all interactions that can modify their number or energy. These are known as diffusion-loss equation, for which an extensive treatment can be found in \cite{Berezinsky:2002nc,Aloisio:2008pp,Aloisio:2010he} for our cases of interest. Here the propagation in the intergalactic magnetic fields (treated for protons for instance in \cite{Aloisio:2004jda}) is neglected, and a general form for the transport equation can be given by:
\begin{equation}
    \frac{dn_i(E,t)}{dt}=-\frac{d}{dE}\left[ \frac{dE(t)}{dt}n_i(E,t)\right]-\frac{n_i(E,t)}{\tau_i(E,t)}+Q_i(E,t)
    \label{eq:transport}
\end{equation}
where $n_i(E,t)$ represents the number of particles of species $i$ per volume and energy, with energy $E$ at the time $t$; the variation of $n_i$ with time is due to energy losses (first term at the right side of the equation), particle losses due to decays with lifetime $\tau_i(E,t)$ (second term) and to an injection rate represented by the third term. In the following, the flux will be also used, which can be defined from the quantity $n$ as: 
\begin{equation}
J(E,z=0) = \frac{c}{4\pi} n(E,z=0)~.
\label{eq:f}
\end{equation}

In the next subsections the computation will be specified to the case of UHECR protons and nuclei, using, if needed, the following notation:
\begin{equation}
\label{eq:beta_def}
-\frac{1}{E}\frac{dE}{dt} = \beta_0\left(E \right),
\end{equation}
from which the quantity
\begin{equation}
\label{eq:b_def}
b_0(E)=-\frac{dE}{dt} = E \beta_0 (E) 
\end{equation}
can be also defined.

\subsection{The case of UHECR protons}\label{sec:protons}
In order to compute the spectrum for the case of protons, one should consider that the second term at the right hand side of Eq.~\ref{eq:transport} is zero. Therefore, one can evolve the particles from the time of observation to the cosmological epoch of generation using the adiabatic energy losses $EH(z)$ and $b$ as:
\begin{equation}
    -\frac{dE}{dt}=EH(z) + (1+z)^{2} b_0[(1+z)E]\, ;
    \label{eq:eneloss}
\end{equation}
the dependence of $b$ and $\beta$ from the redshift is explained in App.~\ref{sec:redshiftdependence}.

Let us first compute the flux from a single source; in order to do so, the evolution of the energy as a function of time/redshift has to be evaluated, and the energy intervals at the epoch of production and detection have to be computed.

\begin{figure}[t]
\centering
\begin{tabular}{cc}
\includegraphics[scale=0.25]{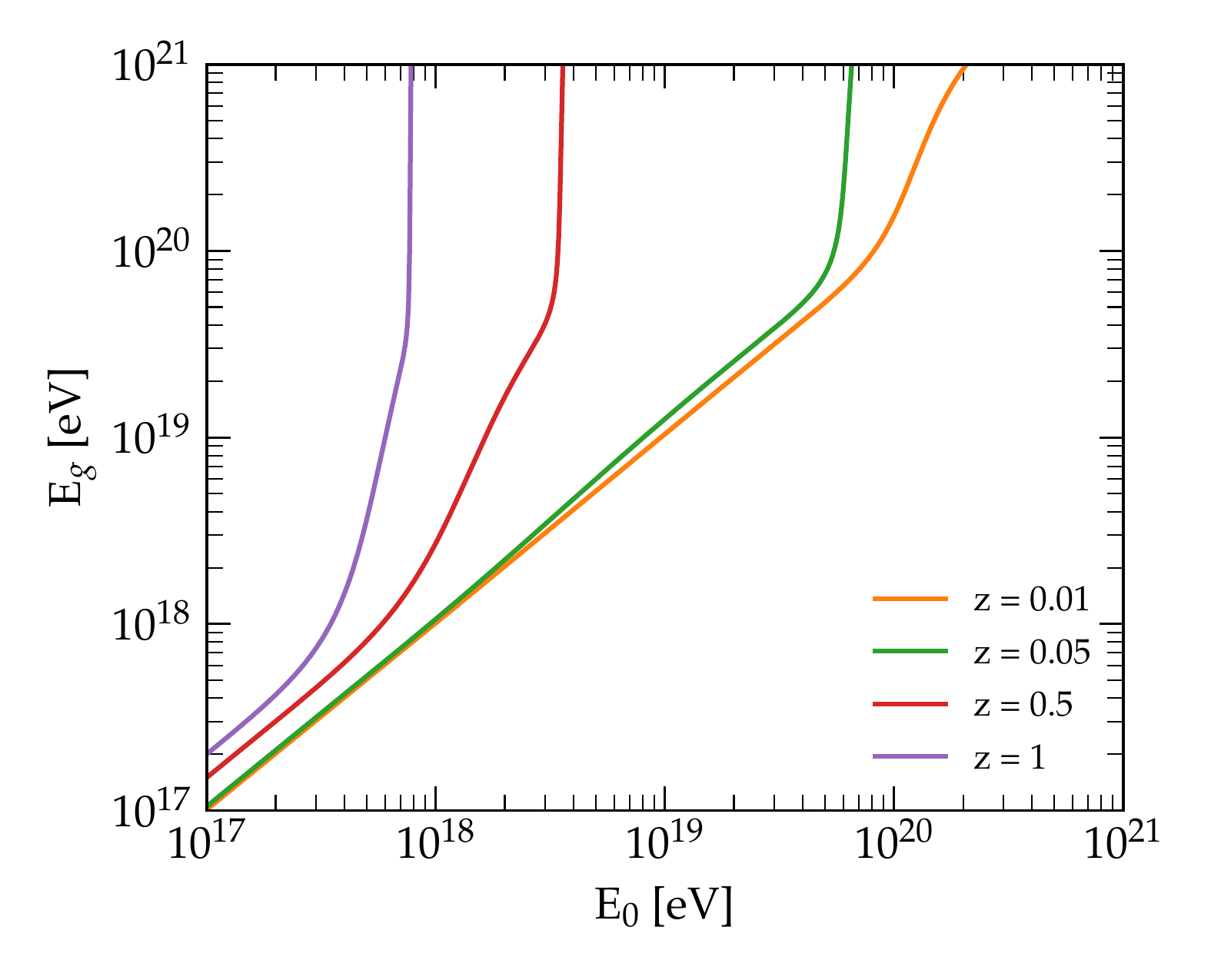}&
\includegraphics[scale=0.25]{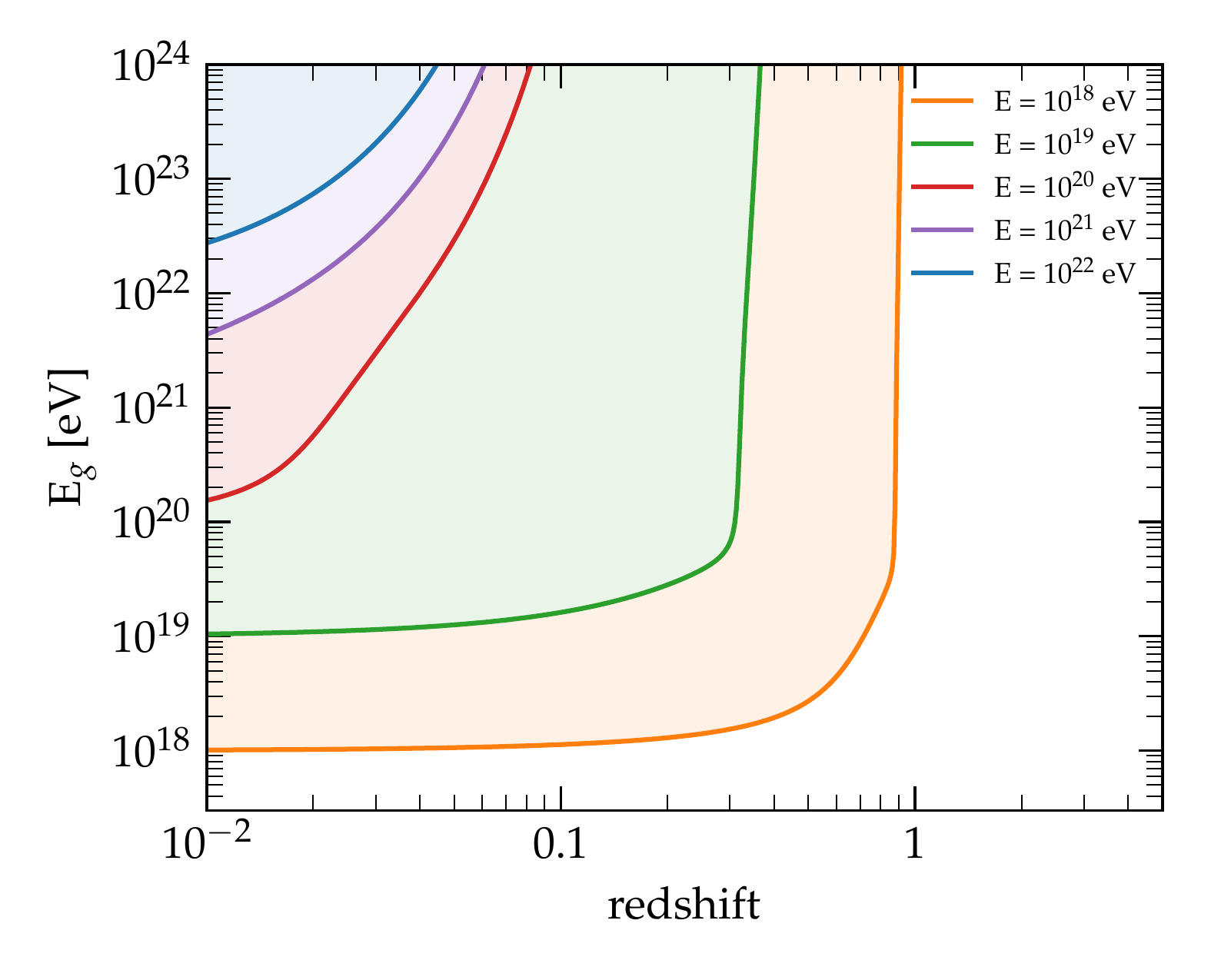}
\end{tabular}
\caption{Left: the energy at generation as a function of the energy at Earth, for different values of the redshift at generation \cite{Sirente}. Right: the energy at generation as a function of the redshift, for different values of the energy at Earth \cite{Sirente}.}
\label{fig:proton_Eg}
\end{figure}

In order to connect the energy intervals, the following energy loss equation has to be solved:
\begin{equation}
\label{energy_loss_p}
\frac{dE_\text{g}}{dt} = - E_\text{g} \beta(E_\text{g},z(t)),
\end{equation}
where the subscript $g$ refers to the energy of the particle at the generation. The $\beta$-function is given by the contribution from interactions and expansion of the Universe
\begin{equation}
\beta(E_\text{g},z(t)) \ \ \rightarrow \ \ \beta(E_\text{g},z(t)) + H(z(t)),
\end{equation}
where $H(z(t))$ is the Hubble parameter at the time $t$. Given the initial condition $E(t=t_0)=E$, where $t=t_0$ is the present time, the solution of Eq.~\ref{energy_loss_p} for a generic time $t<t_0$ is 
\begin{equation}
E_\text{g}(t) = E + \int_t^{t_0} dt' \, E_\text{g}(t')H(z(t')) + \int_t^{t_0} dt' \, E_\text{g}(t')\beta(E_\text{g}(t'),z(t')) .
\end{equation}
This solution can be written using the redshift parameter such that $z(t=t_0)=0$. The energy loss equation for protons can be written as a function of the redshift making use of the transformation (see App.~\ref{sec:cosmology} for more details):
\begin{equation}
\frac{dt}{dz}  = -\frac{1}{H_0 (1+z) \sqrt{(1+z)^{3}\Omega_m+\Omega_{\Lambda}}} 
\label{eq:dtdz}
\end{equation}
as:
\begin{equation}
E_\text{g}(z) = E + \int_0^z d z' \, \frac{E_\text{g}(z')}{1+z'}+ \int_0^z d z' \, \frac{E_\text{g}(z')\beta (E_\text{g}(z'),z')}{H(z')(1+z')}.
\end{equation}
The relations \ref{b_def} and \ref{beta_z} can be used to show that
\begin{equation}
\beta (E,z) = (1+z)^3 \beta_0((1+z)E) = \frac{(1+z)^2}{E}b_0 ((1+z)E), 
\end{equation}
and therefore 
\begin{equation}
\label{energy_loss_z}
E_\text{g}(z) = E + \int_0^z d z' \, \frac{E_\text{g}(z')}{1+z'}+ \int_0^z d z' \, \frac{1+z'}{H(z')}b_0 ((1+z')E_\text{g}(z')).
\end{equation}
If the previous equation is differentiated with respect to $E$, the expansion factor of the energy interval can be computed as 
\begin{equation}
\begin{aligned}
y(z) & = 1+\int_0^z d z' \, \frac{y(z')}{1+z'} + \int_0^z d z' \, \frac{1+z'}{H(z')} \frac{db_0((1+z')E_\text{g}(z'))}{dE} \\
  & = 1+\int_0^z d z' \, \frac{y(z')}{1+z'} + \int_0^z d z' \, \frac{(1+z')^2}{H(z')} \frac{db_0((1+z')E_\text{g}(z'))}{d ((1+z')E_\text{g}(z'))} y(z'),
\end{aligned}  
\end{equation}
whose corresponding differential equation is
\begin{equation}
\label{eq_y}
\frac{1}{y}\frac{dy}{dz} = \frac{1}{1+z} + \frac{(1+z)^2}{H(z)} \frac{db((1+z)E_\text{g}(z))}{d ((1+z)E_\text{g}(z))}.
\end{equation}
The solution of Eq.~\ref{eq_y} is the connection between the energy intervals: 
\begin{equation}
\label{y_sol}
y(z) = (1+z)\exp \left[\frac{1}{H_0}\int_0^z dz' \, \frac{(1+z')^2}{\sqrt{(1+z')^3 \Omega_m + \Omega_\Lambda}} \frac{db_0((1+z')E_\text{g}(z'))}{d ((1+z')E_\text{g}(z'))} \right]\, ,
\end{equation}
which therefore enters in the computation of the expected number of particles at Earth. A complete derivation of the reported calculation can be found in \cite{Berezinsky:2002nc}. Fig.~\ref{fig:proton_Eg} shows the energy at the source as a function of the energy at Earth, corresponding to different redshifts (left) and the energy at the source as a function of the redshift, corresponding to different energies at Earth (right).
\\

The flux from a single source at cosmological distance $z$ will be:
\begin{equation}
\label{flux_singles}
J(E,z) = \frac{1}{(4\pi)^2}\frac{Q(E_\text{g}(E,z),z)}{(1+z_\text{g})\chi^2}\frac{dE_\text{g}}{dE}
\end{equation}
where the connection between the energy intervals appears; $\chi$ is the comoving radial coordinate (as defined in App.~\ref{sec:cosmology}) and $Q$ is the generation rate per unit energy, that can be expressed as
\begin{equation}
Q(E_{\mathrm{g}})=Q_0 \left(\frac{E_{\mathrm{g}}}{E_{0}}\right)^{-\gamma} f_{\mathrm{cut}}(E_{\mathrm{g}})\, .
\label{eq:Qsource}
\end{equation}
being $\gamma$ the spectral index and $f_{\mathrm{cut}}(E_{\mathrm{g}})$ a function that describes the cut-off of the flux at the source, which might depend on the acceleration process and/or the interactions suffered by the particles in the source environment; the normalization factor $Q_0$ will be explained later. 
The calculation can be extended for considering a distribution of sources as:
\begin{equation}
\label{flux_distrs}
J(E) = \frac{1}{(4\pi)^2} \int dV  \frac{\widetilde{Q}(E_\text{g}(E,z),z)}{(1+z)\chi^2}\frac{dE_\text{g}}{dE};
\end{equation}
here the generation rate per comoving volume is used, being defined as $\widetilde{Q}=n_0 Q$ where $n_0$ is the number of sources per unit volume. The previous expression can be written as an integral over the redshift as:
\begin{equation}
\label{eq:flux_distrs_dt}
J(E) = \frac{c}{4\pi} \int dz \left|\frac{dt}{dz} \right|  \widetilde{Q}(E_\text{g}(E,z),z)\frac{dE_\text{g}}{dE}
\end{equation}
where the following transformation has been used:
\begin{equation}
\label{dVdz}
\frac{1}{4\pi} \frac{dV}{dz} = (1+z)^3 c d^2_\text{A} \left|\frac{dt}{dz} \right|,  
\end{equation}
being $d_\text{A}$ the angular-diameter distance, as defined in App.~\ref{sec:cosmology}.\\

\begin{figure}[t]
\centering
\includegraphics[scale=0.6]{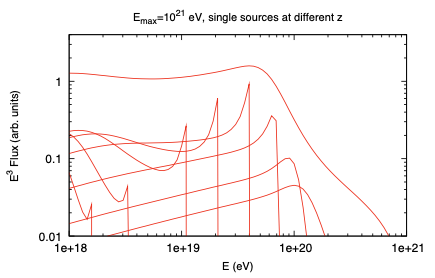}
\caption{Expected flux of cosmic-ray protons at Earth, multiplied by $E^3$, corresponding to protons injected with a power law with $\gamma = 2.6$ and maximum acceleration energy at the source $E_{\mathrm{cut,g}} = 10^{21}$ eV (indicated as $E_{\mathrm{max}}$ in the top of the figure), from a uniform distribution of identical sources (upper line). Expected fluxes from single sources are also shown, with redshift respectively (with decreasing energy cutoff): 0.005 ($\sim$20 Mpc), 0.01 ($\sim 40$ Mpc), 0.03 ($\sim$125 Mpc), 0.1 ($\sim$420 Mpc), 0.2 ($\sim$820 Mpc), 0.3 ($\sim$1200 Mpc), 0.5 ($\sim$1890 Mpc), 0.7 ($\sim$2500 Mpc), 0.9 ($\sim$3000 Mpc).
From \cite{Boncioli_PhDThesis2011}.}
\label{fig:single_diffuse_flux}
\end{figure}

\begin{figure}[t]
\centering
\begin{tabular}{cc}
\hspace{-0.8cm}
\includegraphics[scale=0.47]{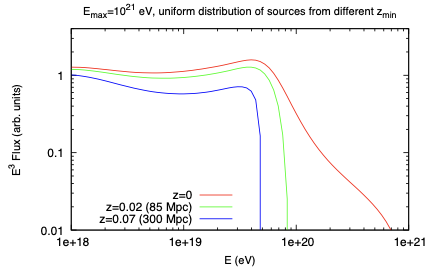}&
\hspace{-0.3cm}
\includegraphics[scale=0.44]{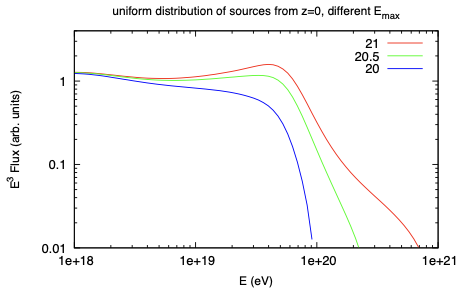}
\end{tabular}
\caption{Expected flux of cosmic-ray protons at Earth, multiplied by $E^3$, corresponding to protons injected with a power law with $\gamma = 2.6$ and maximum acceleration energy at the source $E_{\mathrm{cut,g}} = 10^{21}$ eV (indicated as $E_{\mathrm{max}}$ in the top of the figure), from a uniform distribution of identical sources. Left: $E_{\mathrm{cut,g}} = 10^{21}$ eV at injection, uniform distribution of sources starting from different $z_{\mathrm{min}}$. Right: $E_{\mathrm{cut,g}}$ (indicated as $E_{\mathrm{max}}$ in the top of the figure) variable, uniform distribution of sources starting from $z_{\mathrm{min}}=0$. From \cite{Boncioli_PhDThesis2011}.}
\label{fig:protons_flux}
\end{figure}

Therefore the expected flux of UHECR protons at Earth can be computed as in Eq.~\ref{eq:flux_distrs_dt}, corresponding to an expanding Universe homogeneously filled by sources of accelerated primary UHE protons with some choice for the spectrum at the source reported in Eq.~\ref{eq:Qsource}. An example is reported in Fig.~\ref{fig:single_diffuse_flux}, where the contribution of single sources at different distances and the cumulative spectrum (multiplied by $E^{3}$) are depicted, corresponding to $\gamma=2.6$ and $E_{\mathrm{cut}}=10^{21}$ eV, being defined as  $f_{\mathrm{cut}}(E_{\mathrm{g}})=\exp(-E_{\mathrm{g}}/E_{\mathrm{cut,g}})$. The closest sources show no deviations from the initial spectrum, while corresponding to increasing distances a bump is visible, as expected due to the rapid pile-up of the protons below the photo-pion threshold \cite{Berezinsky:1988wi}. The abrupt suppression of the individual spectra, which is also reflected in the diffuse spectrum at the highest energies, is the effect of the energy losses due to photo-pion processes, as predicted in \cite{Greisen:1966jv,Zatsepin:1966jv}, commonly known as "GZK" suppression from the initials of the authors. The bump is then expected to be smoother in the diffuse flux, because individual peaks are located at different energies. Below the bump, a dip is visible at larger distances, as expected due to the pair production energy losses \cite{Berezinsky:2005cq}. In the diffuse flux the protons in the dip should be collected from a large volume, thus one could expect this feature to be less dependent on the distribution of sources. The measured change of the slope at $5\times 10^{18}$ eV, the ankle, could therefore, in the context of pure proton composition of UHECRs, be interpreted as a signature of the propagation of the protons through the CMB.\\
Attributing the suppression of the spectrum to the GZK effect is however not entirely justified. In fact, at the highest energies the visible Universe in terms of cosmic rays is strongly dependent on the local distribution of sources. As an example, in Fig.~\ref{fig:protons_flux} (left panel) the change in the suppression at the highest energies as due to the redshift of the closest source is shown: the farther is the closest source, the lower is the energy of the suppression. A similar effect can be obtained if the maximum energy at the acceleration is varied (see Fig.~\ref{fig:protons_flux}, right panel), indicating that the shape of the suppression is degenerate in terms of these variations, which would contribute to the depletion of the flux as well as the "pure" GZK effect.

\subsection{The case of UHECR nuclei}\label{sec:nuclei}
In the following, the chain of equations describing the nuclear cascade in the extragalactic space is discussed, using the Lorentz factor instead of the energy, as done in \cite{Aloisio:2008pp,Aloisio:2010he}. The primary nucleus loses energy due to interactions with background photons and can photo-disintegrate; the generated secondary nucleus $A$ is then considered as produced again homogeneously in the space with a rate $Q_{A}(\Gamma, z)$ depending on the solution of the transport equation for $A_0$ corresponding to the current $(\Gamma,z)$. To simplify the treatment, nuclei are assumed to suffer only photo-disintegrations with the emission of one nucleon. Under this simple hypothesis, the injection rate of the secondary nuclei is:
\begin{equation}
Q_A(\Gamma,z)=\frac{n_{A_0}(\Gamma,z)}{\tau_{A_0}(\Gamma,z)}
\label{eq.inj_sec}
\end{equation}
where $n_{A_0}(\Gamma,z)$ is the equilibrium distribution of the parent nucleus and $\tau_{A_0}(\Gamma,z)$ is the photo-disintegration life-time of $A_0$. The resulting equation chain is then:
\begin{eqnarray}
{\frac{\partial n_{A_0}(\Gamma,t)}{\partial t} }- \frac{\partial}{\partial \Gamma}[n_{A_0}(\Gamma,t){b_{A_0}(\Gamma,t)}]+
{\frac{n_{A_0}(\Gamma,t)}{\tau_{A_0}(\Gamma,t)}} & = &{Q_{A_0}(\Gamma,t)}\\ \nonumber
{\frac{\partial n_{A_0-1}(\Gamma,t)}{\partial t} }- \frac{\partial}{\partial \Gamma}[n_{A_0-1}(\Gamma,t){b_{A_0-1}(\Gamma,t)}]+
{\frac{n_{A_0-1}(\Gamma,t)}{\tau_{A_0-1}(\Gamma,t)}}& = &{\frac{n_{A_0}(\Gamma,t)}{\tau_{A_0}(\Gamma,t)}}\\ \nonumber
& \vdots &\\
{\frac{\partial n_A(\Gamma,t)}{\partial t}} - \frac{\partial}{\partial \Gamma}[n_A(\Gamma,t){b_A(\Gamma,t)}]+
{\frac{n_A(\Gamma,t)}{\tau(\Gamma,t)}} & = &{\frac{n_{A+1}(\Gamma,t)}{\tau_{A+1}(\Gamma,t)}}~;
\label{eq.transeq}
\end{eqnarray}
here the notation of Eq.~\ref{eq:transport} has been used, with $\Gamma$ instead of $E$, as motivated by the conservation of the Lorentz factor in the photo-disintegration processes.\\

Here the solution for the secondary nuclei is derived, from which the solutions for primary nuclei and secondary nucleons can be easily obtained.\\
The characteristic equation for the transport reads
\begin{equation}
\frac{d\Gamma(t)}{dt} = -b_A(\Gamma,t).
\label{eq.char}
\end{equation}
With $\Gamma(t)$ taken on the characteristics, the term $b_A(\Gamma,t)\frac{\partial n_A(\Gamma,t)}{\partial \Gamma}$ disappears and Eq.~\ref{eq.transeq} takes the form
\begin{equation}
{\frac{\partial n_A(\Gamma,t)}{\partial t}} + n_A(\Gamma,t) \left[ -\frac{\partial b^A_{pair}(\Gamma,t)}{\partial \Gamma} -\frac{\partial b^A_{ad}(\Gamma,t)}{\partial \Gamma} + \tau^{-1}_A(\Gamma,t) \right] = Q_{A}(\Gamma,t);
\label{eq.transeq2}
\end{equation}
with this choice the time $t$ becomes the only variable. The resulting solution is then
\begin{equation}
n_A(\Gamma,t) = \int_{t_g}^t dt^{'} Q_A(\Gamma,t^{'}) \exp \left[ -\int^t_{t^{'}} dt^{''}\left( -P_1(\Gamma,t^{''}) -P_2(\Gamma,t^{''}) + \tau^{-1}_A(\Gamma,t^{''})   \right)  \right]
\label{eq.n_A(t)}
\end{equation}
where the notation used is 
\begin{eqnarray}
P_1(\Gamma,z) & = &\frac{\partial b^A_{ad}(z)}{\partial \Gamma} = H(z)\\
P_2(\gamma,z) & = &\frac{\partial b^A_{pair}(\Gamma,z)}{\partial \Gamma} = \frac{Z^2}{A} (1+z)^3 \left( \frac{\partial b^p_0(\Gamma^{'})}{\partial \Gamma^{'}}  \right)_{\Gamma^{'}=(1+z)\Gamma}
\label{eq.notation}
\end{eqnarray}
that are written in terms of $z$ instead of time $t$, as can be done using the relation \ref{eq:dtdz}. The solution can then be written at $z=0$ as 
\begin{eqnarray}
& n_A(\Gamma,z=0) = \int_{0}^{z_{\mathrm{max}}} dz^{'} \frac{Q_A(\Gamma^{'}(\Gamma,z^{'}))}{(1+z^{'})H(z^{'})} \times & \\ 
& \exp \left[ \int_{0}^{z'}  dz^{''}  \frac{P_1(\Gamma,z^{''})}{(1+z^{''})H(z^{''})}  \right]
\exp \left[ \int_{0}^{z'}  dz^{''}  \frac{P_2(\Gamma,z^{''})}{(1+z^{''})H(z^{''})}  \right]
\exp \left[ - \int^{t_0}_{t^{'}}  \frac{dt^{''}}{\tau_A(\Gamma,t^{''})}  \right] ~.&
\label{eq.n_A(z)}
\end{eqnarray}
The last integration is kept over time to show that it represents the suppression factor for the survival time of the nucleus $A$, that can be read also as 
\begin{equation}
\eta(\Gamma^{'},z^{'}) = \int^{t_0}_{t^{'}}  \frac{dt^{''}}{\tau_A(\Gamma,t^{''})} = \int^{z^{'}}_z dz^{''}  \frac{\tau^{-1}_A(\Gamma^{''},z^{''})}{(1+z^{''})H(z^{''})} ~;
\label{eq.eta}
\end{equation}
the product of the first two exponents of Eq.~\ref{eq.n_A(z)} gives the ratio of energy intervals calculated in the previous section. Finally,
\begin{equation}
n_A(\Gamma,z) = \int_{z}^{\infty} dz^{'} \frac{Q_A(\Gamma^{'}(\Gamma,z^{'}))}{(1+z^{'})H(z^{'})} \frac{d \Gamma^{'}}{d \Gamma} e^{-\eta(\Gamma^{'},z^{'})},
\label{eq.n_Afin} 
\end{equation}
gives the density of a species $A$ at a given $(\Gamma,z)$; to compute the density at Earth, the lower bound of the integral must be set to $z=0$.\\

The density of primary nuclei can be computed taking into account that  the injection term in Eq.~\ref{eq.n_Afin} is proportional to $\Gamma^{-\gamma_{g}}$; instead in the case of secondary nucleons the transport equation is obviously devoid of the photo-disintegration term:
\begin{equation}
\frac{\partial n_n(\Gamma,t)}{\partial t} - \frac{\partial}{\partial \Gamma}[n_n(\Gamma,t){b_n(\Gamma,t)}] = {\frac{n_{A+1}(\Gamma,t)}{\tau_{A+1}(\Gamma,t)}}~,
\label{eq.transeqp}
\end{equation}
where the injection term is the same as in the case of secondary nuclei, because of the hypothesis of conservation of the Lorentz factor in the photo-disintegration process. The solution $n_n(\Gamma,t)$ is then similar to Eq.~\ref{eq.n_Afin}, without the suppression factor for the survival time of the nucleus $A$. A complete derivation of these quantities is given in \cite{Aloisio:2008pp}.\\
The number of particles per volume and energy $n$ for each species can be finally converted to the flux as done in Eq.~\ref{eq:f}.\\

The solution of a set of coupled kinetic equations described in this section can be implemented semi-analytically, both for the case of the CMB and EBL. However, Monte Carlo approaches are also extensively used for the computation of the UHECR propagation, as for instance done in \cite{Aloisio:2017iyh,AlvesBatista:2022vem}. Typically this approach consists in the evaluation of the probability of a particle to survive an interaction; for instance, in the case of a photo-disintegration, taking into account  Eq.~\ref{eq:intl_general} for the interaction rate, the probability reads:
\begin{equation}
  P_A(\Gamma,z) = \exp  \left( - \int \frac{1}{\tau_A(\Gamma(z'),z')} \left| \frac{dt}{dz'} \right| dz' \right)\, ;   
\end{equation}
this is usually evaluated in steps in redshift, and the energy losses are computed separately. With the Monte Carlo method it is therefore also possible to compute the interaction point of the particle, that is in turn the generation point of the secondary particle (which is also fundamental in order to compute the fluxes of secondary particles such as neutrinos or gamma rays).\\

\begin{figure}[t]
\centering
\includegraphics[scale=0.2]{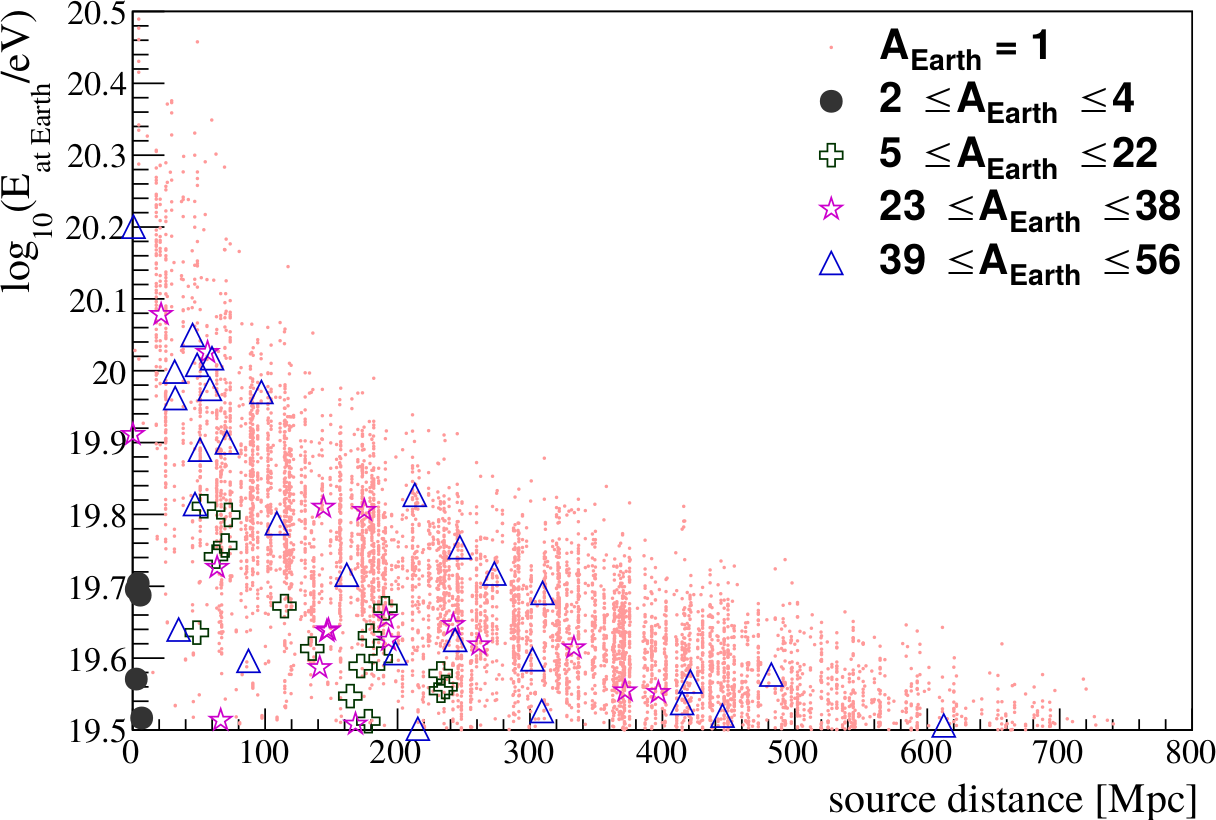}
\caption{Energy at Earth as a function of the distance at which the cosmic ray is created, for different nuclear species, computed from a \textit{SimProp} 2.4 \cite{Aloisio:2017iyh} simulation. From \cite{armando}.}
\label{fig:horizon}
\end{figure}

It is interesting to note here that, similarly to the pion production for protons, also the photo-disintegration entails the disappearance of nuclei of a certain species, because of the creation of lighter fragments, and the excitation of the GDR for the interactions with CMB photons happens at similar energies as for the threshold of the pion production of protons on CMB. In addition, the energy loss lengths for the photo-disintegration processes are of similar order of magnitude as the one for the pion production from protons. For these reasons, the visible Universe in terms of cosmic rays at the highest energies is similar for protons and heavy nuclei (as one can see in Fig.~\ref{fig:horizon}, where simulations of \textit{SimProp} 2.4 \cite{Aloisio:2017iyh} are used to show the energy at Earth as a function of the distance of the source that produced a cosmic-ray particle of the species indicated in the legend). This implies that the interpretation of the suppression of the spectrum, experimentally observed at the highest energies, is also degenerate in terms of the chemical composition of the cosmic rays, in addition to the other possible motivations due to the pion production effect (if protons), the distribution of the sources and the maximum energy at the acceleration, as shown in Sec.~\ref{sec:protons}. 

\begin{figure}[t]
\centering
\begin{tabular}{cc}
\hspace{-0.8cm}
\includegraphics[scale=0.35]{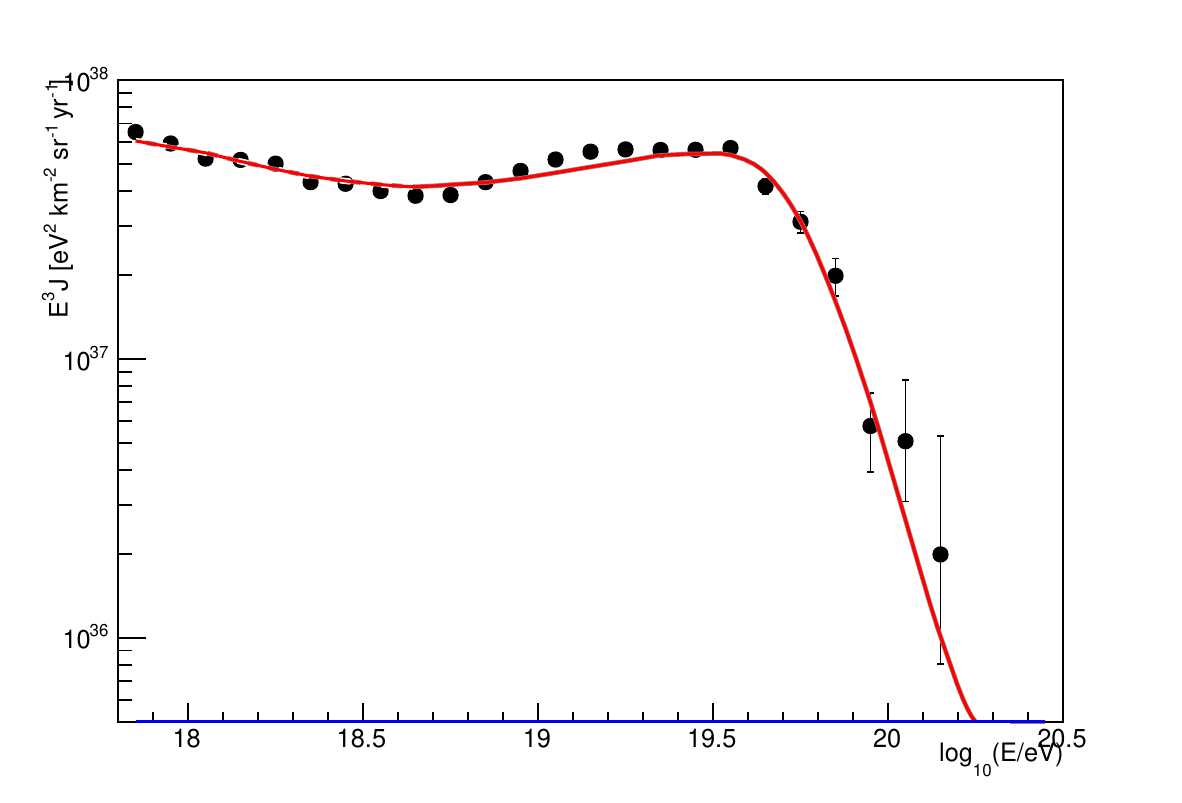}&
\hspace{-1.7cm}
\includegraphics[scale=0.35]{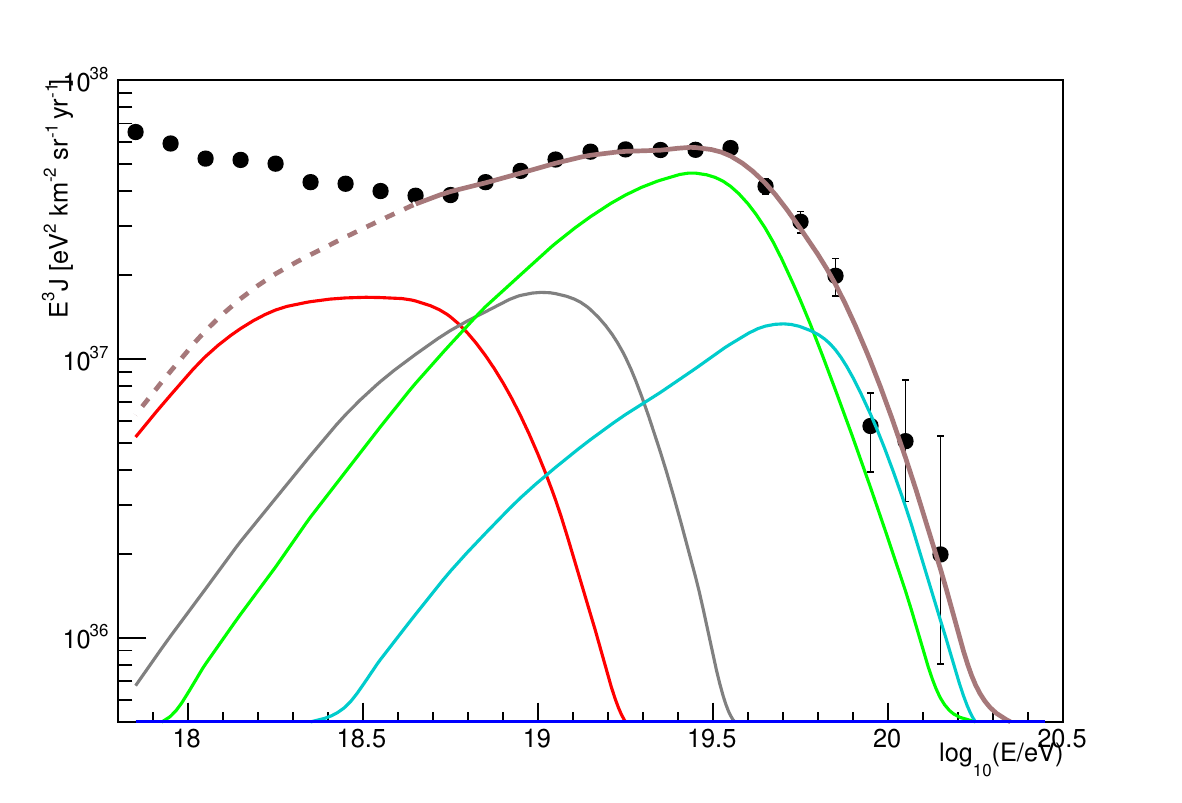}
\end{tabular}\\
\hspace{-1cm}
\includegraphics[scale=0.65]{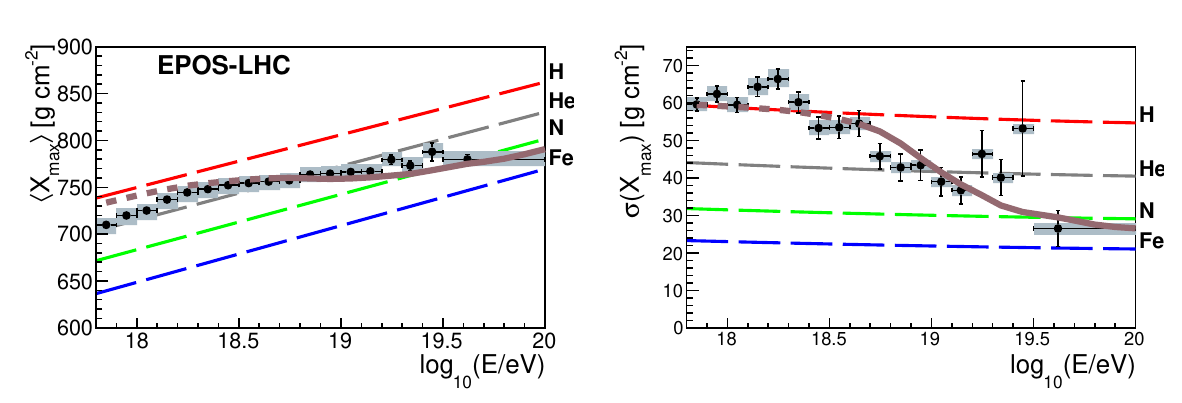}
\caption{Upper panels: Simulated energy spectrum of UHECRs (multiplied by $E^3$ at the top of the Earth atmosphere, obtained with a pure-proton composition with $E^{-2.4}$, $R_{\mathrm{max}}=19.7$ V and $m=3.5$ at the sources (left panel) and with a mixed mass composition (dominated by CNO) with $E^{-1}$, $R_{\mathrm{max}}=18.7$ V and $m=0$ (the parameters are expressed at the sources, following Eq.~\ref{eq:eq1bis}). Partial spectra in the left panel are grouped as: $A=1$ (red), $2\le A \le4$ (grey), $5\le A \le22$ (green), $23 \le A \le28$ (cyan), while the total (all-particle) spectrum is shown in brown. Lower panels: Average and standard deviation of the $X_{\mathrm{max}}$ distribution as expected from the astrophysical scenario corresponding to the propagated flux reported in the upper right plot (brown line) together with pure hydrogen (red), helium (grey), nitrogen (green) and iron (blue) lines corresponding to the predictions of the hadronic interaction model EPOS-LHC \cite{Pierog:2013ria}. The data points are from \cite{Veberic:2017hwu}.\\
The scenario reported in the upper right and lower panels is  re-adapted from the best fit found in \cite{PierreAuger:2016use}. The simulations used in these plots are obtained with \textit{SimProp} 2.4 \cite{Aloisio:2017iyh}.}
\label{fig:AugerCF}
\end{figure}

Understanding the origin of the suppression of the spectrum, as well as of its other features, requires considering other CR observables such as the ones connected to the chemical composition, as proposed for instance in \cite{PierreAuger:2016use}, and constitutes one of the main open issues in cosmic-ray astrophysics. In \cite{PierreAuger:2016use}, a uniform distribution of identical sources emitting cosmic-ray nuclei with a power law of energy up to some maximum rigidity is assumed, following 
\begin{equation}
    \widetilde{Q}_{A}(E) 
    = \widetilde{Q}_{0A} \cdot \left( \frac{E}{E_0} \right)^{-\gamma}
    \cdot \begin{cases}
        1, & E \leq Z_A \cdot R_{\mathrm{cut}}; \\
        \exp\left(1-\frac{E}{Z_A \cdot R_{\mathrm{cut}}}\right), & E > Z_A \cdot R_{\mathrm{cut}},
    \end{cases}\label{eq:eq1bis}
\end{equation}
and the propagation in the extragalactic space is taken into account in order to produce the expected energy spectrum and mass composition at Earth. In Eq.~\ref{eq:eq1bis}, $\widetilde{Q}_{0A}$ accounts for the percentage of a certain nuclear species at the source (evaluated at a fixed energy $E_0$). In Ref. \cite{PierreAuger:2016use}, the expected fluxes are then fitted to the data of the Pierre Auger Observatory, so that the corresponding spectral details and the nuclear species at the source are determined. The best-fit results for energies above the ankle predict an intermediate mass composition (dominated by the carbon-nitrogen-oxygen group), and a hard injection spectrum ($\sim E^{-1}$) and low rigidity ($\sim 10^{18.7}$ V, corresponding to a maximum energy of $\sim 10^{18.7}$ eV for protons and $\sim 10^{20.15}$ eV for silicon nuclei, the heaviest nuclear species found in this study) at the escape from the sources. The results are reported in the upper right panel of Fig.~\ref{fig:AugerCF} for the energy spectrum (all-particle and mass groups at Earth) and in the lower panels of Fig.~\ref{fig:AugerCF} for the mass composition observables (the mean value - lower left panel - and the width - lower right panel - of the distributions of the position in the atmosphere at which the shower reaches its maximum number of particles\footnote{The detection of UHECRs cannot be pursued with techniques measuring directly the particles reaching the Earth from the outer space, due to the diminishing intensity of the flux with increasing energy. What is done is instead to take advantage of the Extensive Air Shower (EAS), meaning the cascade of particles initiated in the atmosphere by the incoming UHECR particle. From the simple model for the development of electromagnetic cascades \cite{Heitler}, that connects the maximum number of particles in the shower to the depth in the atmosphere at which radiative losses equal ionization losses (corresponding to $E^{\mathrm{em}}_{\mathrm{c}}$, happening at depth $ X_{\mathrm{max}}$), and enables the determination of the energy $E$ of the primary particle from
\begin{equation}
     X_{\mathrm{max}} \propto  \ln (E/E^{\mathrm{em}}_{\mathrm{c}}) \, ,
\end{equation}
a generalization can be drawn in order to model also showers initiated by hadrons \cite{Matthews:2005sd}. 
The nuclear composition of the cosmic rays is in reality determined on a statistical basis from the longitudinal development of the shower, if the primary nucleus of mass $A$ and energy $E$ is treated as a superposition of $A$ nucleons of energy $E' = E/A$ (\textit{superposition model}). Using the superposition model, the depth at which there is the maximum number of particles in a shower initiated by a nucleus of mass $A$ can be obtained taking into account the one of protons as:
\begin{equation}
    \langle X^{\mathrm{A}}_{\mathrm{max}}  \rangle = \langle X^{\mathrm{p}}_{\mathrm{max}}  \rangle - D_{\mathrm{p}} \ln A
    \label{eq:xmaxA}
    \end{equation}  
where $D_{\mathrm{p}}$ is the elongation rate for protons, defined as $D_{\mathrm{p}}=\mathrm{d} \langle X^{\mathrm{p}}_{\mathrm{max}}  \rangle/ \mathrm{d} \ln E = 25$ $\mathrm{g \,cm^{-2}}$. See also \cite{Kampert:2012mx} for a review on mass composition measurements.\\

}). These results can be compared to the case reported in the upper left panel of Fig.~\ref{fig:AugerCF}, where the spectrum of pure protons with $E^{-2.4}$ up to a maximum rigidity of $\sim 10^{19.7}$ V (with the same shape of the cutoff function at the source as in Eq.~\ref{eq:eq1bis}) is reported (for the proton case, a cosmological evolution of the sources as $m=3.5$ has been used, while for the mixed composition $m=0$; the cosmological evolution is included in the injection spectrum in Eq.~\ref{eq.n_Afin} as $(1+z)^m$). In both the pure-proton and  mixed-composition case, the suppression at the highest energies cannot be due to the sole effect of the extragalactic propagation; in fact, as suggested by \cite{PierreAuger:2022atd}, the low value of the maximum rigidity found at the sources could be also responsible of the measured suppression. Regarding the ankle feature, while in the case of the mixed-composition it cannot be reproduced with a single population of sources, and an additional component at lower energies should be taken into account \cite{PierreAuger:2022atd}, the case of pure protons seems to reproduce the spectrum features of the suppression and of the ankle with the chosen parameters, but cannot account for the behavior of the mass composition observables and is reported here only for illustration purposes.\\

The study of CR interactions in the extragalactic space, as treated in these lectures, is fundamental to contribute to the understanding of the characteristics of the main observables of cosmic rays. In addition, the same interactions here discussed could be taken into account in the modeling of the source environment; together with the diffuse processes in the source site, these interactions could be responsible of the reprocessing of the cosmic rays after the acceleration process and influence the shape of the spectra at the escape in the region of the ankle \cite{Unger:2015laa}. Models that take into account both interactions in astrophysical candidate sources of cosmic rays and in extragalactic space could therefore enhance our capability of explaining the features of the UHECR energy spectrum at Earth, taking into account the UHECR mass composition.

\section{Cosmogenic neutrinos and photons}\label{sec:neutrinos}
The interactions suffered by cosmic ray particles during their passage through CMB and EBL generate secondary particles that in turn decay. The photo-pion processes involve the excitation of the Delta resonance, whose de-excitation can produce charged or neutral pions, that in turn decay and produce neutrinos or gamma-rays, called \textit{cosmogenic}.\\

\begin{figure}[t]
\centering
\hspace{-1.cm}\includegraphics[scale=0.38]{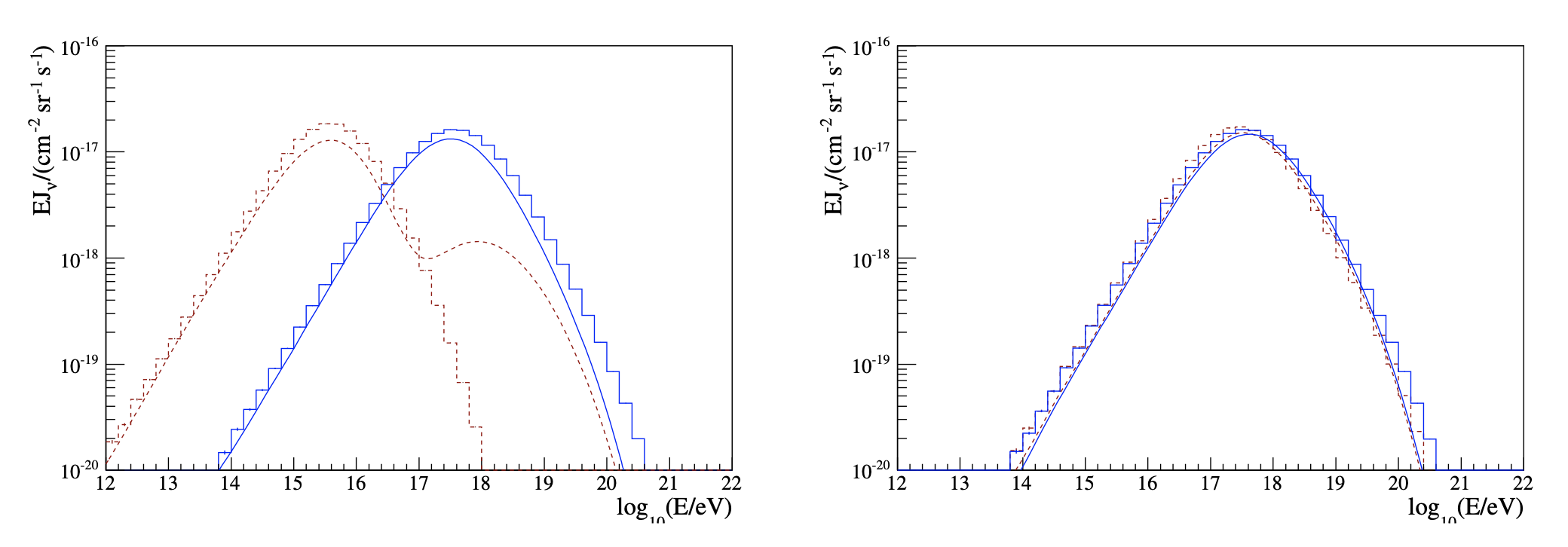}
\caption{Left panel: fluxes of electron (blue solid line) and anti-electron (red dashed line) neutrinos generated in propagation of protons through CMB. Right panel: fluxes of muon (blue solid line) and anti-muon (red dashed line) neutrinos. The histograms are obtained from \textit{SimProp} simulations, while the lines are taken from \cite{Engel:2001hd}. Reproduced with permission from \cite{Aloisio:2015ega}.}
\label{fig:neu_JCAP2015}
\end{figure}

Let us first discuss the case of the production of charged pions as:
\begin{equation}
\begin{aligned}
    p+\gamma_{\mathrm{bkg}} \rightarrow \Delta^{+} \rightarrow\, & \pi^{+} + n\\
    & \pi^{+} \rightarrow  \mu^{+}+\nu_{\mu}\\
    & \mu^{+} \rightarrow e^{+}+\nu_{\mathrm{e}} + \bar{\nu}_{\mu}
\end{aligned}\label{eq:pgammachain}
\end{equation}
Therefore three neutrinos (with flavor composition of $\nu_{\mathrm{e}}:\nu_{\mu}:\nu_{\tau}=1:2:0$) for each pion are produced. From considerations about the inelasticity, as reported in Sec.~\ref{sec:rate}, the pions carry about 20\% of the energy of the primary proton.
The expected flux of cosmogenic neutrinos at Earth depends on the characteristics of the spectrum of protons emitted from the sources, as described in \cite{Lipari:2007su}. From Eq.~\ref{eq:pgammachain} one can see that the expected fluxes of electron/muon neutrinos and muon anti-neutrinos (reported in Fig.~\ref{fig:neu_JCAP2015}, right panel) are expected to be of equal intensity (they are produced for each $p\gamma$ interaction), and peaked at the same energy (they carry on average 5\% of the energy of the initial proton). A contribution of electron anti-neutrinos can arise from the decay of neutrons, and its flux is expected to be peaked at lower energies. Anti-electron neutrinos which can be produced from the decay chain of negative pions (possibly produced in multi-pion productions) can also contribute to the high-energy neutrino peak (see Fig.~\ref{fig:neu_JCAP2015}, left panel).

The expected neutrino flux is also connected to the photon fields with which the protons can interact. In order to trigger a photo-pion production off a CMB photon (average energy $\epsilon \approx 7\cdot 10^{-4}$ eV), a more energetic proton with respect to the case of the EBL field is needed, being the energy of the photon in the nucleus rest frame $\epsilon'\approx \Gamma \epsilon$. As a consequence, the high energy peak of the neutrino flux is expected to be originated from interactions off CMB, while the low energy peak from interactions off EBL. 

\begin{figure}[t]
\centering
\begin{tabular}{cc}
\hspace{-0.5cm}
\includegraphics[scale=0.35]{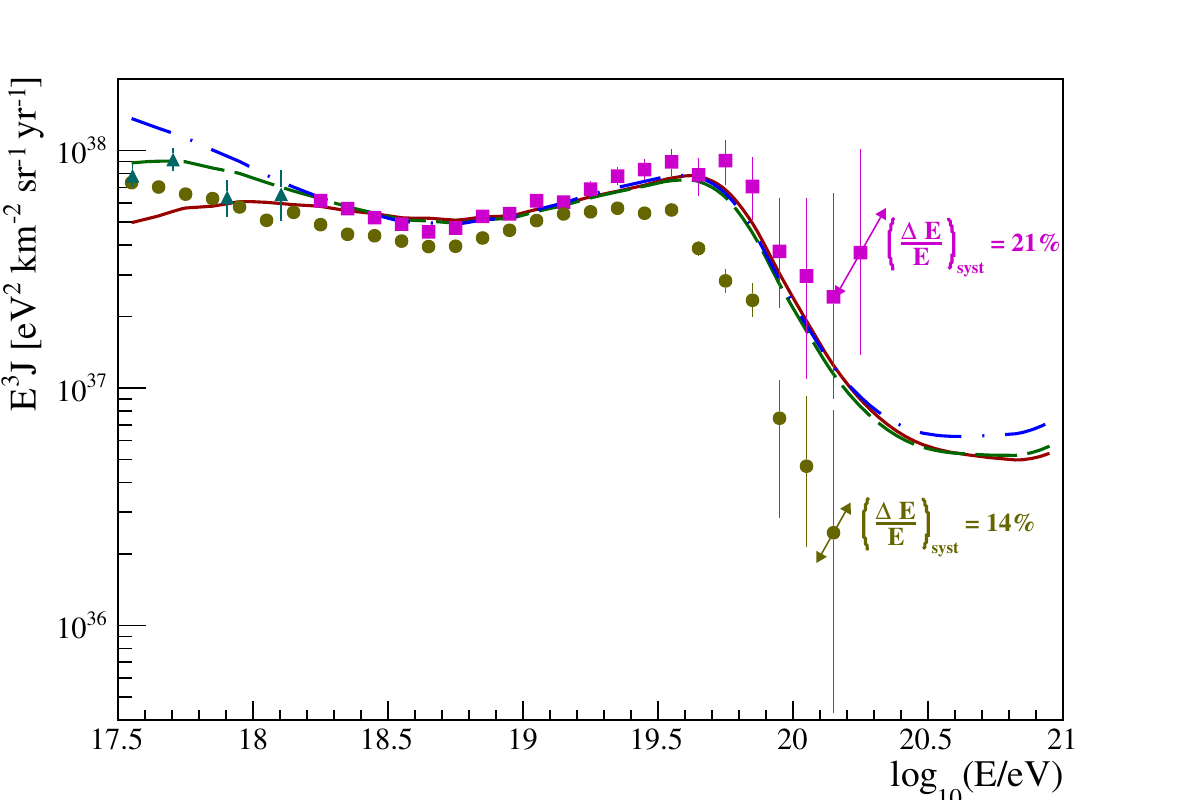}&
\hspace{-0.4cm}\includegraphics[scale=0.35]{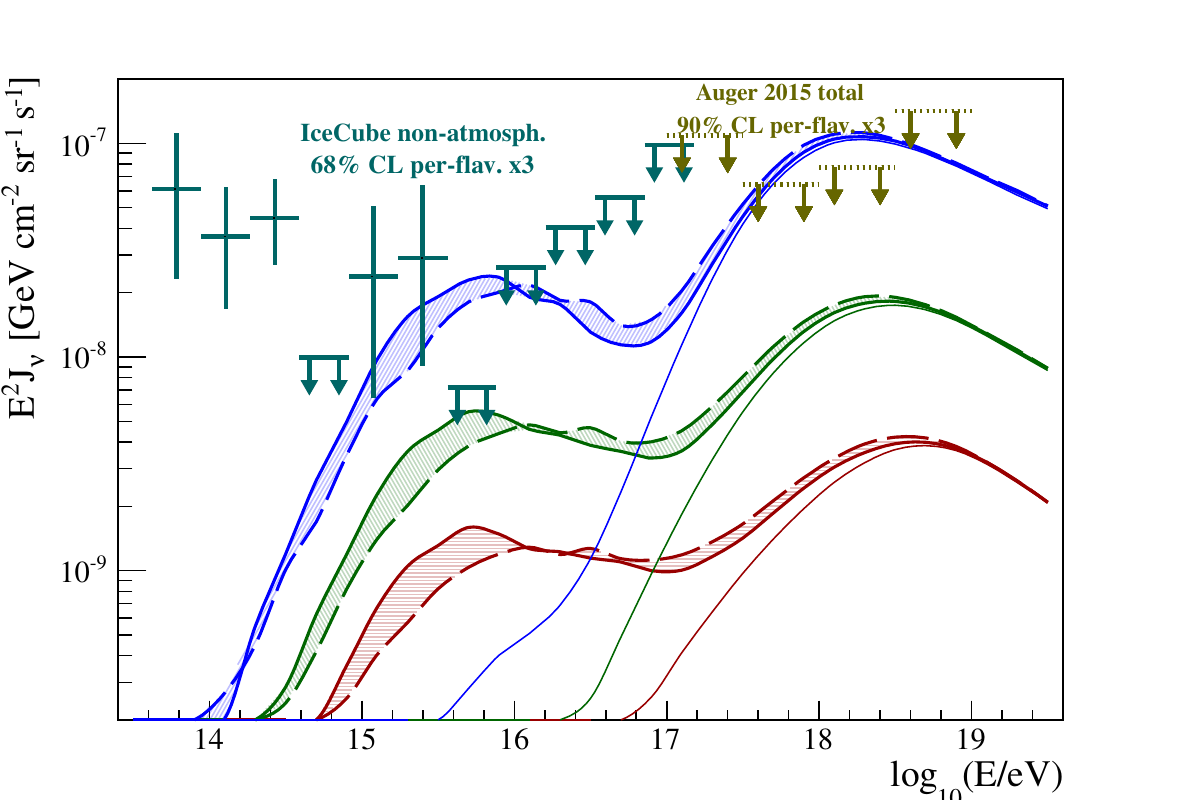}
\end{tabular}
\caption{Left panel: cosmic-ray fluxes expected at Earth (in this case, the propagation through both CMB and EBL is computed) in scenarios with pure proton composition, with various models for the cosmological evolution of sources (solid red line: no evolution; dashed green line: SFR evolution \cite{Hopkins:2006bw}; dot-dashed blue line: AGN evolution \cite{Gelmini:2011kg}). For comparison also the experimental data from the Telescope Array \cite{TelescopeArray:2013bxp} and the Pierre Auger Observatory \cite{PierreAuger:2013qtf} are shown (magenta and olive green dots, respectively). Right panel: fluxes of neutrinos in the same scenarios. Reproduced with permission from \cite{Aloisio:2015ega}.}
\label{fig:neu_flux_simprop}
\end{figure}

Neutrinos can travel for long distances unimpeded; this is the reason why they can be accumulated for a large portion of the Universe, as can be seen in the different colors of the lines in Fig.~\ref{fig:neu_flux_simprop} (right panel: here the case of pure-proton composition for the cosmic rays that generated the neutrinos is taken into account). Here the cases of no cosmological evolution of the distribution of UHECR sources is reported (red line) together with the case of Star Forming Rate (SFR) evolution (green line) \cite{Hopkins:2006bw} and the case of evolution of the high-luminosity Active Galactic Nuclei (AGN) as suggested in \cite{Gelmini:2011kg}. The effect of the cosmological evolution, to be considered in Eq.~\ref{eq:flux_distrs_dt} or in Eq.~\ref{eq.n_Afin} (depending on the mass of the CR) as a term like $(1+z)^{m}$ entering in the distribution of sources, is expected to be more important while increasing the redshift (if $m>0$). Due to the interactions of UHECR particles, the effect of the cosmological evolution of sources is less dominant than for neutrinos, as can be seen in the left panel of Fig.~\ref{fig:neu_flux_simprop}, where the propagated spectra of protons are different only below $10^{18.2}$ eV.\\ 

Combining the information from UHECRs and neutrinos can be, therefore, very relevant to the aim of constraining the cosmological distribution of the sources, which remains undetermined if only UHECRs are taken into account. Examples of  combined studies involving both the UHECR spectrum (with pure proton composition) and the expected cosmogenic neutrinos can be found for instance in \cite{Heinze:2015hhp,PierreAuger:2019ens}.
The expected flux of cosmogenic neutrinos is strongly related to the characteristics of the flux of UHECRs at the escape from the sources, as well as from the details of the cosmological evolution of UHECR sources. The maximum energy of UHECRs determines the cutoff of the neutrino flux, while the shape of the neutrino flux is mainly dependent on the spectral index of the UHECR spectrum. The chemical composition of UHECRs is also affecting the expected neutrino flux; due to the fact that the photo-pion production is a process involving the nucleons in the nucleus, it is convenient to compute the value of the threshold Lorentz factor, which for the photo-pion production reads $\Gamma_{\mathrm{th}}\approx 7\times10^{10}$. Therefore the energy threshold for a photo-pion process for a generic nucleus is $E_{\mathrm{th}}=A \Gamma^{p}_{\mathrm{th}}m_p c^{2}$; nuclei heavier than hydrogen would then require $A$ times the energy of a proton in order to excite the resonance responsible for this process. For this reason, scenarios involving heavier nuclear composition of UHECRs predict a smaller neutrino flux \cite{Heinze:2019jou,AlvesBatista:2018zui}. It is important to stress here that the fits of the UHECR energy spectrum and composition in terms of astrophysical scenarios usually involve only the energy range above the ankle; if the entire UHE range is taken into account, as done for instance in \cite{PierreAuger:2022atd}, a proton component, with a soft energy spectrum at the sources, would be required in order to reproduce the energy region below the ankle, together with a heavier contribution, probably coming from the final spectrum of the Galactic cosmic rays. This proton component would be connected to a more efficient neutrino production. In particular, if the light CR component below the ankle is connected to the interactions happening in the source environment, as for instance reported in \cite{Unger:2015laa}, neutrinos produced in the same reactions in the sources are expected. These could be produced both in interactions of UHECRs with the photon fields in the source, which can be treated as the processes described for the extragalactic propagation where the extragalactic photon field is substituted by the photon field typical of the source, as well as by interactions of UHECRs with the matter in the source environment, as for instance treated in \cite{Condorelli:2022vfa} for the case of the nuclei of starburst galaxies. Details of the production of neutrinos in these interactions can be found in the lecture notes by P. Serpico at this School \cite{Serpico}.\\

\begin{figure}[t]
\centering
\includegraphics[scale=1.3]{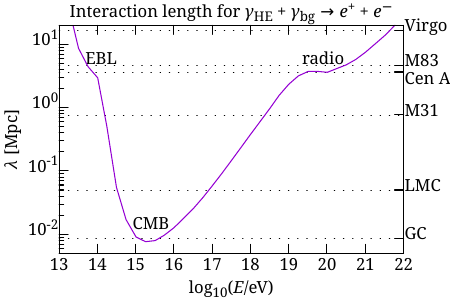}
\caption{Interaction length of gamma rays for pair production in several background fields. Courtesy of A. di Matteo.}
\label{fig:gammaray_horizon}
\end{figure}

Neutral pions can be produced in the de-excitation of $\Delta^{+}$, giving origin to cosmogenic photons:
\begin{equation}
\begin{aligned}
    p+\gamma_{\mathrm{bkg}} \rightarrow \Delta^{+} \rightarrow & \pi^{0} + p\\
    & \pi^{0} \rightarrow \gamma + \gamma
    \end{aligned}\label{eq:cosm_gamma}
\end{equation}
Similarly to cosmic rays, cosmogenic photons can interact with photon backgrounds (interaction lengths for these processes are shown in Fig.~\ref{fig:gammaray_horizon}), giving rise to electromagnetic cascades, where in turn high-energy photons can be absorbed due to pair production, and electrons undergo inverse Compton and produce synchrotron radiation, transferring energy to the range below $10^{14}$ eV.\\

\begin{figure}[t]
\centering
\includegraphics[scale=0.5]{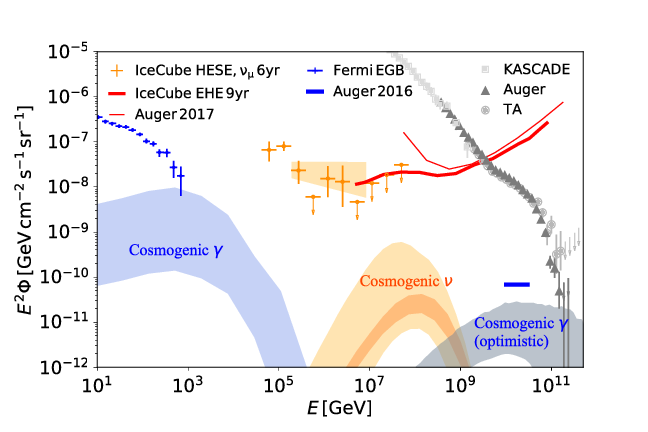}
\caption{Cosmogenic photon (blue) and neutrino (orange) fluxes for UHECR models that fit the Pierre Auger Observatory data including spectrum and composition. Reproduced with permission from \cite{AlvesBatista:2019tlv}.}
\label{fig:multim}
\end{figure}

It is interesting to notice that the energy densities in UHECRs, high-energy neutrinos and gamma-rays are similar (as reported in Fig.~\ref{fig:multim}). This would suggest that the origin of these different messengers is strongly connected and support a multi-messenger view, which might strongly improve our understanding of the characteristics of UHECRs and other messengers.

\appendix
\section{Interaction rate}\label{sec:interactionrate}
The expression for the interaction length (Eq.~\ref{eq:intl_general}) can be derived from fundamental quantities of the theory of interactions. We report here a procedure similar to what done in \cite{Gaisser2,Dermer}. The relativistic rate of interactions per unit volume can be defined as the number of collisions per unit volume and time, between particles 1 and 2 with masses $m_1$ and $m_2$, with densities $n_1$ and $n_2$ in the reference frame $K$, as
\begin{equation}
\dot{n} = \dfrac{dN}{dV dt}\, .
\label{eq:collisionrate}
\end{equation}
Due to relativistic length contraction, computing the densities in the frame $K$ with respect to the one where the particles are at rest (indicated with the superscript "0") will give
\begin{equation}
n_i=\Gamma_i n^0_i
\label{eq:rate}
\end{equation}
 with $i=1,2$; the Lorentz factor $\Gamma_i$ is in this simplified case considered to be the same for all particles of the same type. If we apply then the transformation to the frame where particles of species 2 are at rest, we find 
 \begin{equation}
\dot{n}= c \beta_r \sigma n^0_2 n'_1
\label{eq:rate_2}
\end{equation}
where $\sigma$ is the cross section of the process and $c\beta_r$ is the relative speed of particles of type 1 in the rest frame of particles of type 2, and can be expressed as:
\begin{equation}
\beta_r = \left(\frac{(p_1^{\mu} p_2^{\mu})^{2}-m^{2}_1 m^{2}_2}{(p_1^{\mu} p_2^{\mu})^{2}}\right)^{1/2}\, ; 
\label{eq:beta_r}
\end{equation}
$\beta_r=1$ in case at least one of the particles is a photon. The cross section is a function of the relative Lorentz factor of a particle in the rest frame of the other one, which reads:
\begin{equation}
\Gamma_r = \frac{p_1^{\mu} p_2^{\mu}}{m^{2}_1 m^{2}_2} = \Gamma_1 \Gamma_2(1-\vec{\beta}_1 \cdot \vec{\beta}_2)\, .
\label{eq:gamma_r}
\end{equation}
Let us consider the density of particles of species 1 in the rest frame of particles of species 2, $n'_1=\Gamma_r n^0_1=\Gamma_r n_1/\Gamma_1$, and $n^0_2 = n_2/\Gamma_2$, that therefore implies:
\begin{equation}
\dot{n}= c \beta_r \sigma(\Gamma_r) (1-\vec{\beta}_1 \cdot \vec{\beta}_2) n_1 n_2
\label{eq:rate_2}
\end{equation}
which is valid if the particles are monoenergetic; in a realistic case, in which particles have a distribution of energy and velocity (also in space) the reaction rate can be written in a more general way as:
\begin{equation}
\dot{n}= \frac{c}{1+\delta} \int \int \beta_r \sigma(\Gamma_r) (1-\vec{\beta}_1 \cdot \vec{\beta}_2) dn_1 dn_2 \, ,
\label{eq:rate_dist}
\end{equation}
with $\delta=0 (1)$ for interactions of different types of particles (self interacting particle distributions).  
It is important to notice that the reaction rate in Eq.~\ref{eq:rate_dist} has been derived in the proper frame of particles of species 2, but due to the fact that $\dot{n}$ is an invariant quantity, it also gives the reaction rate in the frame $K$\footnote{It is an invariant quantity because it is expressed as the ratio of two invariant quantities, the number of collision $dN$ and the quantity $dV\,dt$.}. 

Let us now consider the case of a particle traversing a photon field; in the case where one of the particles is a photon we have $\delta=0$, $\beta_r=\beta_2=1$, and the quantities expressed in $\Gamma_r$ are then expressed in terms of $\epsilon_r = \Gamma_1 \epsilon (1-\beta_1 \cos \theta_{12})$. In addition, it is also useful to consider now the differential spectral density which reads:
\begin{equation}
n(\vec{p})=\frac{dn}{d^{3} \vec{p}}=\frac{dn}{p^{2} dp d\Omega}\, ,
\label{eq:n_p}
\end{equation}
$dn=n(\vec{p})p^{2} dp d\Omega$ and $d\Omega=d\cos\theta \,d\phi$. Then the generic interaction rate is:
\begin{equation}
\dot{n}= \frac{c}{1+\delta} \int d\Omega_1 \int dp_1 \, p_1^{2}\,  n_1(\vec{p}_1) \int d\Omega_2 \int dp_2\, p_2^{2}\, n_2(\vec{p}_2)\, \beta_r\, \sigma(\Gamma_r)\, (1-\beta_1  \beta_2 \cos\psi) \, ,
\label{eq:rate_generic}
\end{equation}
where $\psi$ is the angle between the directions of the interacting particles or photons.
For our case study, we want to consider the rate of interactions of particles in a photon field. For a single particle we have $n_1(\vec{p}_1)=n_1\delta(p-p_1)\delta(\cos\theta_1 -1)\delta(\phi_1)/(4\pi p^{2}_1)$ and therefore we can write:
\begin{equation}
\dot{N}(p_1)=\frac{\dot{n}}{n_1} = c\int^{2\pi}_0 d\phi \int^{1}_{-1} d\cos\theta (1-\beta_1 \cos\theta) \int^{\infty}_0 d\epsilon\,  n_{\mathrm{ph}}(\epsilon, \Omega) \sigma(\epsilon_r)
\end{equation}
where $n_{\mathrm{ph}}=dN/dVd\epsilon d\Omega$ is the photon distribution function and $\epsilon_r$ is the invariant energy of the event, that coincides with the photon energy in the frame where the particle is at rest. Taking into account the Lorentz transformation in the reference frame of the particle $1$, the photon energy $\epsilon'=\epsilon_r$ is then given by
\begin{equation}
\epsilon'=\Gamma_1\epsilon (1-\beta_1 \cos \theta),
\end{equation}
and the corresponding differential is
\begin{equation}
d\epsilon' = -\Gamma_1 \epsilon \beta_1\,d\cos \theta,
\end{equation}
where $\Gamma_a$ is the Lorentz factor of the particle 1. Performing the integration over the azimuth angle, we obtain the expression of the interaction rate as
\begin{equation}
\dot{N}(p_1)=\dfrac{dN_\text{int}}{dt} = \dfrac{c}{2\Gamma_1^2}\int_{\epsilon'_\text{th}}^\infty \sigma(\epsilon')\epsilon' \int_{\epsilon'/2\Gamma_1}^\infty \dfrac{n_{\mathrm{ph}} (\epsilon)}{\epsilon^2} \,d\epsilon\,d\epsilon'.
\end{equation}
where $\epsilon'_\text{th}$ is the energy threshold for the reaction in the reference frame of the particle.
\section{Cosmology}\label{sec:cosmology}
Being the CR sources located at cosmological distances, the Robertson-Walker metric describing a homogeneous and isotropic expanding Universe has to be used \cite{Dermer,KolbTurner}, and can be written as a function of the comoving radial coordinate $\chi=r/R(t)$, being $r$ the space coordinate and $R(t)$ the dimensionless scale factor:
\begin{equation}
    ds^{2} = -c^{2} dt^{2} + R^{2}(t)\left(\frac{d\chi^{2}}{1-k \chi^{2}} + \chi^{2}d\Omega^{2} \right)\, .
\label{eq:RWmetric}    
\end{equation}
Objects with fixed $\chi$ in the universe are separated by a distance that is determined by the variation of the scale factor $R(t)$. The $k$ parameter can be chosen to be $> 0$ ($\Omega > 1$), $< 0$ ($\Omega < 1$) or $= 0$ ($\Omega  = 1$) for spaces of constant positive, negative or zero spatial curvature, respectively. In the following, we will always refer to $\Omega = \Omega_m + \Omega_{\Lambda}$, then $k = 0$ ($\Omega$ is here representing the density).
From the definition of the metric, one can compute the time required for the propagation of the photon from its emission (starred frame) to the detection in an expanding space-time as:
\begin{equation}
    c\int^{t}_{t_*} \frac{d t'}{R(t')}= -\int^{0}_r \frac{d\chi}{\sqrt{1-k\chi^{2}}}\, ;
\end{equation}
if we consider another event, detected at a later time, and we also take into account he fact that $R(t)$ does not change if the events are separated by a small interval of time, we can write the previous integral as: 
\begin{equation}
   \frac{\Delta t}{R(t)} = \frac{\Delta t_*}{R(t_*)}, \,\,\, \Delta t = \frac{R}{R_*} \Delta t_* = (1+z) \Delta t_*\, .
\end{equation}
Taking into account the definition of the redshift $z$ from the relativistic Doppler effect, which reads
\begin{equation}
   z = \frac{\lambda-\lambda_*}{\lambda_*} = \frac{\nu_*}{\nu} -1 
\end{equation}
we can therefore write the scale factor (and the other relevant quantities) as a function of $z$:
\begin{equation}
\frac{R}{R_*}=1+z,\,\,\, \frac{\nu_*}{\nu}=\frac{\epsilon_*}{\epsilon}=\frac{\Delta t}{\Delta t_*}=1+z\, .
\label{eq:redshift}
\end{equation}

The Hubble relation, which can be written as $\vec{v}(\vec{r})=H(t)\vec{r}$, can be also shown in terms of the scale factor as:
\begin{equation}
    H(t)=\frac{1}{\vec{r}}\frac{d\vec{r}}{dt}=\frac{1}{R(t)}\frac{dR(t)}{dt}\, ,
    \label{eq:HR}
\end{equation}
where we have used the definition $\vec{r}=R(t)\chi \hat{r}$. The Hubble constant at time $t_*$ is $H(t_*)=\dot{R}(t_*)/R(t_*)$. The current estimate of the Hubble constant at the present epoch, together with the corresponding time and distance are:
\begin{equation}
H_0\approx 69 \mathrm{\frac{km}{s}\frac{1}{Mpc}}, \,\,\, t_{\mathrm{H}}=\frac{1}{H_0}\approx 14.2 \times 10^{9}\, \mathrm{yr}, \,\,\,   D_{\mathrm{H}}=\frac{c}{H_0}\approx 4350 \, \mathrm{Mpc}.
\end{equation}
while the estimates of the other cosmological parameters are:
\begin{equation}
    \Omega_m \approx 0.3, \,\,\, \Omega_{\Lambda} \approx 0.7 \, .
\end{equation}
Writing the scale factor as a function of the redshift $R(z)=R_0/(1+z)$, therefore $R^{-1}(dR/dz)=-1/(1+z)$ and $H=\dot{R}/R=R^{-1}(dR/dz)(dz/dt)=-(dz/dt)/(1+z)$ and we obtain the conversion factor between the time and the redshift as:
\begin{equation}
\frac{dt}{dz}  = -\frac{1}{(1+z)H(z)} = -\frac{1}{H_0 (1+z) \sqrt{(1+z)^{3}\Omega_m+\Omega_{\Lambda}}}\, ;
\label{eq:time_redshift}
\end{equation}
in the last expression we have used the redshift dependence of $H(z)$ as can be derived by the Friedman equation and considering also the density contributions \cite{KolbTurner}.\\

We report here the definition of distances can be useful in the computation of the CR fluxes at Earth \cite{Dermer}. Let us define the \textit{proper distance} as the one between two objects that would be measured at the same time $t$; at the present time, this is given by the comoving coordinate (the distance between the two objects stays constant over time, if they only move with the Hubble flow and do not have peculiar motion), therefore we define the \textit{comoving distance} as:
\begin{equation}
    d_c(z)= \chi = ct = c\int^{\chi/c}_0 dt =  c \int^z_0  \left|\frac{dt}{dz'}\right| (1+z') \, dz' = \frac{c}{H_0} \int^z_0 \frac{1}{\sqrt{(1+z')^3\Omega_m+\Omega_{\Lambda}}} \, dz'
    \label{eq:dL}
\end{equation}
and the \textit{light-travel distance} as 
\begin{equation}
    d_T(z)= c \int^z_0  \left|\frac{dt}{dz'}\right| dz' = \frac{c}{H_0} \int^z_0 \frac{1}{1+z'} \frac{1}{\sqrt{(1+z')^3\Omega_m+\Omega_{\Lambda}}} dz'\, .
    \label{eq:dT}
\end{equation}

Let us now consider the photon flux measured from a source with proper distance $d_c$ and radiating energy per time as $\mathcal{L}=d\mathcal{E}/dt$; this is related to the energy flux from a source with isotropic luminosity $\mathcal{L_*}=d\mathcal{E_*}/dt_*$ at \textit{luminosity distance} $d_L$ as
\begin{equation}
    \Phi = \frac{d\mathcal{E}}{dA\, dt}=\frac{d\mathcal{E}/dt}{4\pi d^2_c}=\frac{d\mathcal{E_*}/dt_*}{4\pi d^2_L} \, .
    \label{eq:flux_lum}
\end{equation}
The fluence is defined as the integral of $\Phi$ over a time interval
\begin{equation}
    \mathcal{F}=\int^{t_2}_{t_1} dt \Phi(t)
    \label{eq:fluence}
\end{equation}
and is connected to the apparent isotropic energy as:
\begin{equation}
    \mathcal{E}=\frac{4\pi d^2_L \mathcal{F}}{1+z}\, .
    \label{eq:iso_ene}
\end{equation}
Considering the expansion of the universe, $d\mathcal{E_*}=\epsilon_* dN = \epsilon (1+z) dN = (1+z) d\mathcal{E}$ and $dt_*=dt/(1+z)$. From this we have then $d\mathcal{E}/dt=(1+z)^{-2}d\mathcal{E_*}/dt_*$, meaning that the photons of the source are redshifted by a factor $(1+z)$ and the time dilation increases the time interval between two photon emissions and that of their observations again by $(1+z)$. The luminosity distance can be linked to the comoving one as 
\begin{equation}
    d_L(z)= (1+z) d_c = \frac{c}{H_0}(1+z) \int^z_0 \frac{1}{\sqrt{(1+z')^3\Omega_m+\Omega_{\Lambda}}} \, dz'\, ,
    \label{eq:dL}
\end{equation}
for a flat universe.\\

\begin{figure}[t]
\centering
\includegraphics[scale=0.35]{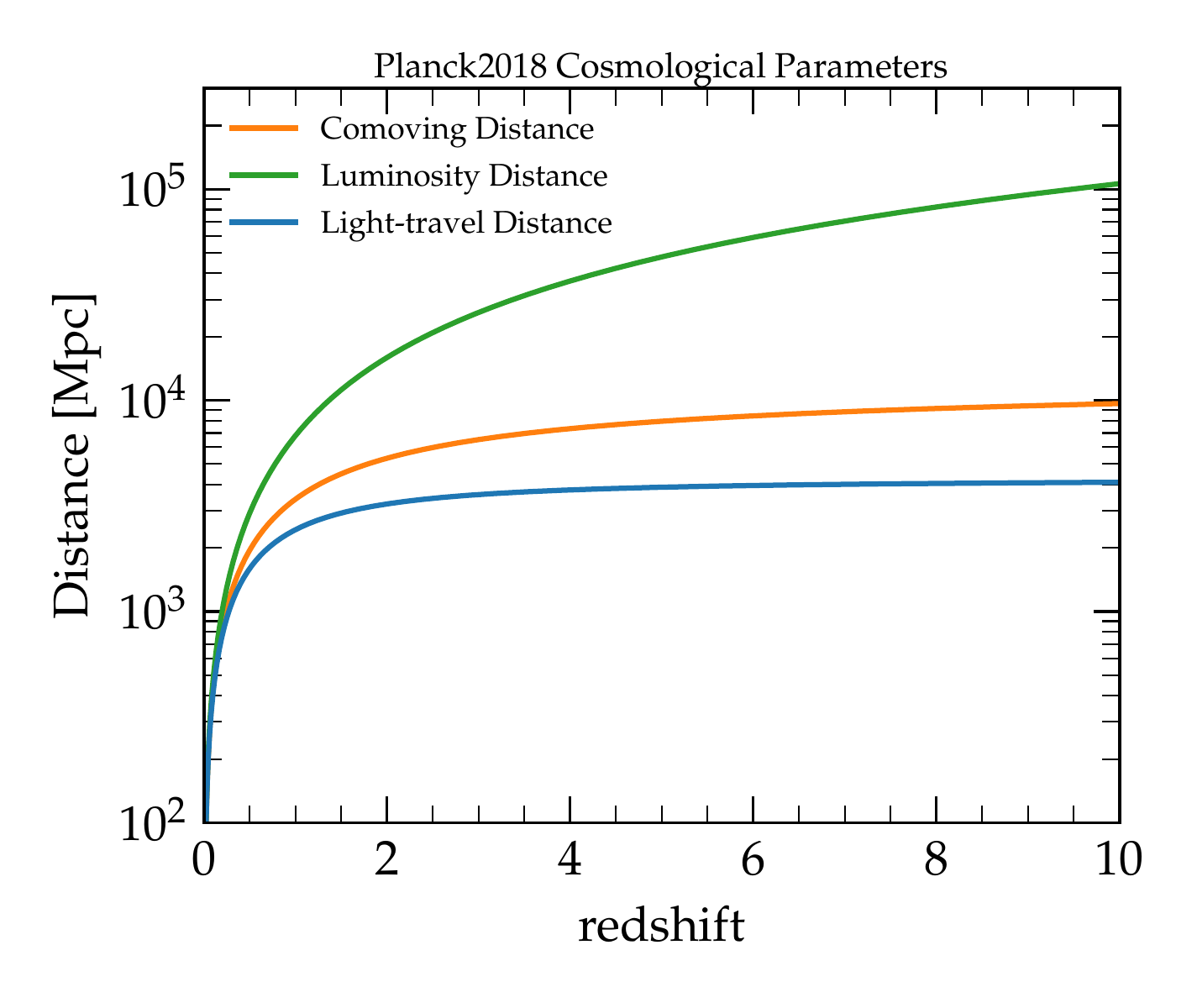}
\caption{Cosmological distances as a function of the redshift, from \cite{Sirente}.}
\label{fig:cosm-dist}
\end{figure}

We can also consider an object at redshift $z$ with transverse proper dimension $D$, with measured angle $\theta$, given in terms of $D$ and the distance $R_*\chi$ referred to the emission time $t_*$. The angular diameter distance is then $d_A=D/\theta=R_*\chi$. If a source at cosmological distance emits photons with isotropic luminosity $\mathcal{L}_*=d\mathcal{E}_*/dt_*$ and considering the definition of luminosity distance, we have
\begin{equation}
    \Phi=\frac{d\mathcal{E}}{dA dt} = \frac{\mathcal{L}_*}{4\pi d^2_L}=\frac{dA}{4\pi d^2_L} \left| \frac{d\mathcal{E}_*}{d\mathcal{E}} \right| \left| \frac{d t_*}{dt}    \right| \frac{d\mathcal{E}}{dA dt} = \frac{(1+z)^2 dA}{4\pi d^2_L} \Phi 
\end{equation}
and therefore $dA=R^2 \chi^2 d\Omega = 4\pi d^2_L/(1+z)^2$. At the time of the emission we have $dA_*=R_*^2 \chi^2 d\Omega_*$ so that $dA_*/dA=1/(1+z)^2$, and therefore we can write the relation between the luminosity distance and the angular diameter distance as 
\begin{equation}
    d_L = \left(\frac{R}{R_*} \right)^2 R_* \chi = (1+z)^2 d_A\,.
\end{equation}
As a summary, we also report the relations with the other distances:
\begin{equation}
    d_c(z) = (1+z) d_A(z) = \frac{d_L(z)}{1+z}\, ;
\end{equation} 
the cosmological distances are also shown in Fig.~\ref{fig:cosm-dist}.
\section{Dependence of relevant quantities on the redshift}\label{sec:redshiftdependence}
\begin{figure}[t]
\centering
\includegraphics[scale=0.35]{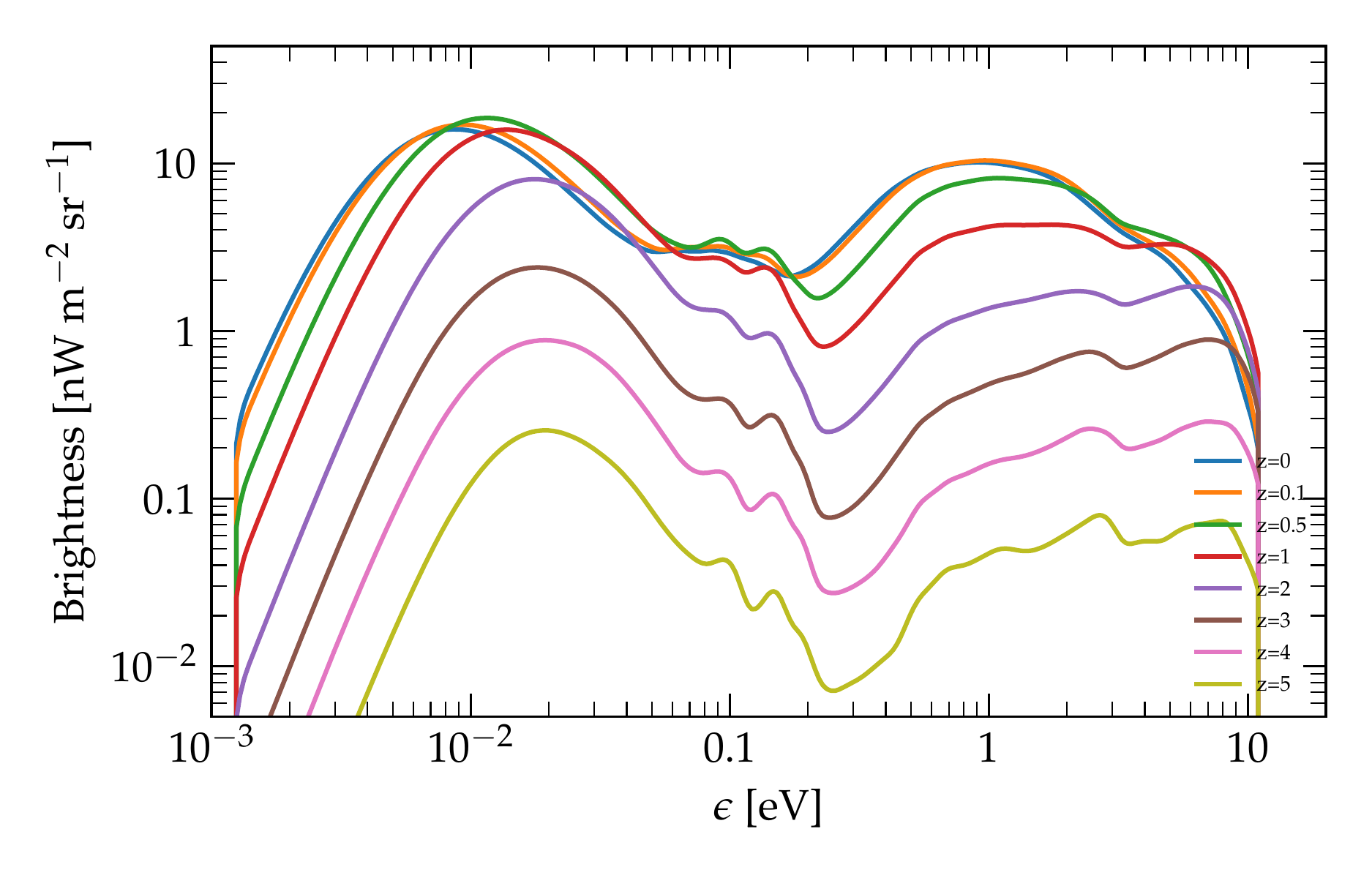}
\caption{The spectral energy distribution of the EBL as modeled in \cite{Saldana-Lopez:2020qzx} as a function of the photon energy, for different values of redshift, and used in \cite{Sirente}.}
\label{fig:EBL_Saldana}
\end{figure}

In order to compute the interaction length of a process at redshift different from zero, how the target photon field appeared in the past has to be known. In Fig.~\ref{fig:EBL_Saldana} the evolution of the EBL intensity as a function of the redsfift is reported, as modeled in \cite{Saldana-Lopez:2020qzx}.

The treatment of the evolution of the CMB as a function of the redshift is instead reported in the following. If we assume that the photon field has been injected in the extragalactic space in the past, the cosmological evolution of its spectral energy density is given by
\begin{equation}
n_{\mathrm{ph}} (\epsilon,z) = (1+z)^2 n_{\mathrm{ph}} \left( \frac{\epsilon}{1+z}\right),
\label{eq:sed_z}
\end{equation}
where $n_{\mathrm{ph}}(\epsilon/(1+z))$ is the spectral energy density at the present time. The factor $(1+z)^2$ comes from the fact that the volume element evolves as $(1+z)^3$, while the energy is redshifted by a factor $(1+z)^{-1}$. Eq.~\ref{eq:sed_z} is valid if there is no feedback from astrophysical sources to the photon field (i.e. the evolution of the field is only driven by the expansion of the Universe). We derive here the evolution of two quantities:
\begin{equation}
n_{\mathrm{ph}} = \int d\epsilon\,n_{\mathrm{ph}} (\epsilon) \  , \ \ \ \ \rho_{\mathrm{ph}} = \int d\epsilon\, \epsilon \, n_{\mathrm{ph}} (\epsilon),
\end{equation}
defined as the number density and the energy density of the photon field, respectively. The cosmological evolution of the number density $n_{\mathrm{ph}} (z)$ is given by
\begin{align}
n_{\mathrm{ph}} (z) = \int d\epsilon\,n_{\mathrm{ph}}(\epsilon,z) &= (1+z)^2 \int d\epsilon\,n_{\mathrm{ph}} \left( \frac{\epsilon}{1+z}\right) \\
& = (1+z)^3 \int d\epsilon\,n_{\mathrm{ph}}(\epsilon) \\
& = (1+z)^3 n_{\mathrm{ph}}.
\label{eq:nphz}
\end{align}
Similarly, one can show that
\begin{equation}
\rho_{\mathrm{ph}} (z) = (1+z)^4 \rho_{\mathrm{ph}}. 
\end{equation}
Starting from Eq.~\ref{eq:intl_general}, we can define the interaction rate of an UHECR with Lorentz factor $\Gamma$ at redshift $z$ as
\begin{align}
\tau^{-1}(\Gamma,z) &= \frac{c}{2\Gamma^2}\int_{\epsilon'_\text{th}}^\infty \sigma(\epsilon')\epsilon' \int_{\epsilon'/2\Gamma}^\infty \frac{n_{\mathrm{ph}} (\epsilon,z)}{\epsilon^2} \,d\epsilon\,d\epsilon' \\
& = \frac{c(1+z)^2}{2\Gamma^2}\int_{\epsilon'_\text{th}}^\infty \sigma(\epsilon')\epsilon' \int_{\epsilon'/2\Gamma}^\infty \frac{n_{\mathrm{ph}} (\epsilon/(1+z))}{\epsilon^2} \,d\epsilon\,d\epsilon'.
\end{align}
We can change the integration variable $\epsilon$ with $\omega(1+z)$; then the interaction rate becomes 
\begin{align}
\tau^{-1}(\Gamma,z) &= \frac{c(1+z)}{2\Gamma^2}\int_{\epsilon'_\text{th}}^\infty \sigma(\epsilon')\epsilon' \int_{\epsilon'/2(1+z)\Gamma}^\infty \frac{n_{\mathrm{ph}} (\omega)}{\omega^2} \,d\omega\,d\epsilon' \\
&= \frac{c(1+z)^3}{2((1+z)\Gamma)^2}\int_{\epsilon'_\text{th}}^\infty \sigma(\epsilon')\epsilon' \int_{\epsilon'/2(1+z)\Gamma}^\infty \frac{n_{\mathrm{ph}} (\omega)}{\omega^2} \,d\omega\,d\epsilon' \\
& = (1+z)^3 \tau^{-1}((1+z)\Gamma,z=0).
\end{align}
The interaction length is given by $l=\tau c$ for UHECRs, then the interaction length at redshift $z$ can be written as 
\begin{equation}
l(E,z) = \frac{l((1+z)E,z=0)}{(1+z)^3},
\end{equation}
where we have used the particle energy instead of the particle Lorentz factor.\\

Here we also derive the dependence on the redshift of other quantities used in the main text. As done in \cite{Berezinsky:2002nc,Berezinsky:1988wi}, we define here
\begin{equation}
\label{beta_def}
-\frac{1}{E}\frac{dE}{dt} = \beta_0\left(E \right),
\end{equation}
from which we can also define the quantity
\begin{equation}
\label{b_def}
b_0(E)=-\frac{dE}{dt} = E \beta_0 (E) ,
\end{equation}
where the subscript 0 refers to the fact that these quantities have been defined at redshift $z=0$. We can include the cosmological evolution of the background photon fields by replacing the photon spectral energy density with $n(\epsilon,t)$. Thus the quantities in Eq.~\ref{beta_def} and \ref{b_def} become 
\begin{equation}
\label{beta_b_def_t}
-\frac{1}{E}\frac{dE}{dt} = \beta \left( E, t \right) 
\end{equation}
and 
\begin{equation}
\label{beta_b_def_t}
b \left( E,t \right) = E \beta \left( E,t \right) .
\end{equation}
Taking into account what has been derived in Eq.~\ref{eq:nphz}, then the relations for $\beta$ and $b$ read
\begin{equation}
\label{beta_z}
\beta (E,z) = (1+z)^3 \beta_0((1+z)E),
\end{equation}
\begin{equation}
\label{b_z}
b (E,z) = (1+z)^2 b_0((1+z)E).
\end{equation}

\acknowledgments
The author acknowledges A. di Matteo and C. Evoli for providing some of the figures shown in these notes. The author acknowledges also S. Rossoni and C. Trimarelli as co-authors of the lecture notes regarding "Ultra-High-Energy Cosmic Rays: Propagation and Detection" \cite{Boncioli:2022ojf} for the lectures given at the first training school of COST Action CA18108 on "Quantum Gravity Phenomenology in the Multi-Messenger Approach" held in Corfu, Greece (September 27th to October 5th 2021), on which some parts of the present notes are based.

\end{document}